\documentclass[pre,twocolumn,twoside,byrevtex,superscriptaddress,floatfix]{revtex4}
\usepackage{color}

\definecolor{todoblue}{RGB}{0, 91, 187}

\usepackage{graphicx,verbatim,enumerate}
\usepackage{amssymb,amsmath}
\usepackage{ifthen}
\newboolean{twocolswitch}
\newboolean{revtexswitch}

\usepackage{mathtools}

\newcommand{\sindex}[1]{}
\newcommand{\nindex}[1]{}

\newcommand{\www}[1]{\url{#1}}
\newcommand{\req}[1]{(\ref{#1})}
\newcommand{\Req}[1]{Eq.~(\ref{#1})}

\newcommand{\dee}[1]{\mbox{d}#1}
\newcommand{\pdiff}[2]{\frac{\partial #1}{\partial #2}}
\newcommand{\pdiffsq}[2]{\frac{\partial^2 #1}{{\partial #2}^2}}
\newcommand{\diff}[2]{\frac{{\rm d}#1}{{\rm d}#2}}
\newcommand{\diffsq}[2]{\frac{{\rm d}^{2}#1}{{\rm d} {#2}^2}}
\newcommand{\tdiff}[2]{\mbox{d} #1/\mbox{d} #2}
\newcommand{\tdiffsq}[2]{\mbox{d}^{2} #1/\mbox{d} {#2}^2}
\newcommand{\tpdiff}[2]{\partial #1/\partial #2}
\newcommand{\tpdiffsq}[2]{\partial^2 #1/\partial {#2}^2}

\newcommand{\kstar}{d^\ast}

\newcommand{\dstar}{d^\ast}
\newcommand{\dstari}{d_i^\ast}

\newcommand{\phifix}{{\phi^{\ast}}}

\newcommand{\phifixb}{{\phi_b^{\ast}}}

\newcommand{\Prob}{\mathbf{P}}

\newcommand{\hfun}{u}
\newcommand{\Hfun}{U}

\newcommand{\dosedistk}{f^{k\star}}

\newcommand{\RtoSprob}{\rho}

\newcommand{\tmax}{t_{\rm max}}

\setboolean{twocolswitch}{true}

\usepackage{multirow}

\begin{document}

\title{
  A Generalized Model of Social and Biological Contagion
}

\author{
\firstname{Peter Sheridan}
\surname{Dodds}
}

\thanks{
  Historical affiliations preserved above.
  Current:
  Vermont Complex Systems Center,
  Computational Story Lab,
  the Vermont Advanced Computing Core,
  Department of Mathematics \& Statistics,
  The University of Vermont,
  Burlington, VT 05401
}

\email{peter.dodds@uvm.edu}

% \homepage{http://smallworld.sociology.columbia.edu/~dodds}

\affiliation{Institute for Social and Economic Research and Policy,
  Columbia University,
  420 West 118th Street,
  New York, NY 10027.}

\author{
  \firstname{Duncan J.}
  \surname{Watts}
  }

\thanks{
Microsoft Research,
641 6th Ave, Floor 7,
New York, NY 10011
}

\email{duncan@microsoft.com}

\affiliation{Department of Sociology,
  Columbia University,
  1180 Amsterdam Avenue,
  New York, NY 10027.}
\affiliation{
        Santa Fe Institute, 
        1399 Hyde Park Road, 
        Santa Fe, NM 87501.}

\date{\today}

\begin{abstract}
  We present a model of contagion that unifies and generalizes existing
models of the spread of social influences and microorganismal infections.
Our model incorporates individual memory of exposure to a
contagious entity (e.g., a rumor or disease), variable magnitudes of
exposure (dose sizes), and heterogeneity in the susceptibility of
individuals.  Through analysis and simulation, we examine in detail
the case where individuals may recover from an infection and then
immediately become susceptible again (analogous to the so-called SIS
model). We identify three basic classes of contagion models which we
call \textit{epidemic threshold}, \textit{vanishing critical mass},
and \textit{critical mass} classes, where each class of
models corresponds to different strategies for prevention or
facilitation.  We find that the conditions for a
particular contagion model to belong to one of the these three classes
depend only on memory length and the probabilities of being
infected by one and two exposures respectively.  These parameters are in principle
measurable for real contagious influences or entities, thus yielding empirical
implications for our model. We also study the case where individuals attain
permanent immunity once recovered, finding that epidemics inevitably die out but
may be surprisingly persistent when individuals possess memory.

  \medskip

  \centering
  \small
  Journal of Theoretical Biology, \textbf{232}, 587--604, 2005.

\end{abstract}

\pacs{89.65.-s,87.19.Xx,87.23.Ge,05.45.-a}
% Social systems, 89.65.-s
% Diseases 87.19.Xx
% Social systems, dynamics of, 87.23.Ge
% Nonlinear dynamics, 05.45.-a

\maketitle

% bifurcation scaling 
% (p-pc)^.5???

% when does the model give only simple class behavior?
% i.e., for what f and g

\section{Introduction}
\label{gcontlong.sec:intro}

Contagion, in its most general sense, is the spreading of an entity
or influence between individuals in a population, via direct or
indirect contact.  Contagion processes
therefore arise broadly in the social and
biological sciences, manifested as, for example, the spread of
infectious diseases~\citep{murray2002,daley1999,anderson1991a,brauer2001,diekmann2000,hethcote2000} and
computer viruses, the diffusion of innovations~\citep{coleman1966,valente1995,rogers2003}, 
political upheavals~\citep{lohmann1994}, and the dissemination of religious
doctrine~\citep{stark1996,montgomery1996}.  
Existing mathematical models of contagion, while motivated in a 
variety of ways depending on the application at hand,
fall into one of only two broad categories, where the critical distinction
between these categories can be explained in terms of the interdependencies between
successive contacts; that is, the extent to which the effect of an
exposure to a contagious agent is determined by the presence or absence
of previous exposures.

The standard assumption in all mathematical models of infectious
disease spreading (for example, the classic SIR model~\citep{kermack1927,murray2002}), 
and also in some models of social contagion~\citep{goffman1964,daley1965,bass1969}, 
is that there is no interdependency between contacts; rather, the infection probability is
assumed to be independent and identical across successive
contacts.  All such models therefore fall into a category that we call 
\textit{independent interaction models}.  
By contrast, what we call \textit{threshold models} assert that an individual can
only become infected when a certain critical number of
exposures has been exceeded, at which point infection becomes highly probable.
The presence of a threshold corresponds to interdependencies of an especially strong
nature: contacts that occur when an individual is near its threshold
are extremely consequential while others have little or no
effect. Threshold models are often used to describe social contagion
(e.g., the spreading of fads or rumors), where individuals either
deterministically~\citep{schelling1973,granovetter1978,watts2002} or
stochastically~\citep{bikhchandani1992,banerjee1992,morris2000,brock2001}
``decide'' whether or not to adopt a certain behavior based in part or in
whole on the previous decisions of others.

An alternative way to think about the interdependence of successive events
is in terms of memory: threshold models implicitly assume the presence
of memory while independent interaction models assume (again implicitly) that the
infection process is memoryless~\footnote{
  Note that independent interaction models typically incorporate a state of immunity,
  which reflects a type of memory, but not the kind with which we are chiefly concerned 
  here---namely, the memory of a contagious influence or entity \textit{prior} to an infection occurring.
}.
Neither class of model, however, is
able to capture the dynamics of contagious processes that possess an
intermediate level of interdependency, or equivalently a variable
emphasis on memory.  Furthermore, the relationship between
interdependent interaction models, threshold models, and any possible
intermediate models is unclear.  Motivated by these
observations, our model seeks to connect threshold and independent interaction models
both conceptually and analytically, and explore the classes of
contagion models that lie between them.  Such an
analysis is clearly relevant to problems of social contagion, in which
memory of past events obviously plays some role, but one that may be
less strong than is assumed by most threshold models.  However, it may
also be relevant to biological disease spreading models, which, to our knowledge,
have previously not questioned the assumption of independence between
successive exposures.  While the independence assumption is indeed
plausible, it has not been demonstrated empirically, and little enough
is understood of the dynamics of immune system response that the
alternative---persistent sub-critical doses of an infectious agent
combining to generate a critical dose---cannot be ruled out.
Furthermore, memory effects are known to be inherent to certain kinds of 
immune system responses, such as allergic response.  Hence if, as we indeed show, only a
slight departure from complete independence is required to alter the
corresponding collective dynamics, then our approach may also shed light
on the spread of infectious diseases.

% 1c. overview and description of model 

\section{Descripton of Model}
\label{gcontlong.sec:model}

\begin{figure*}[tbp]
  \centering
  \includegraphics[width=\textwidth]{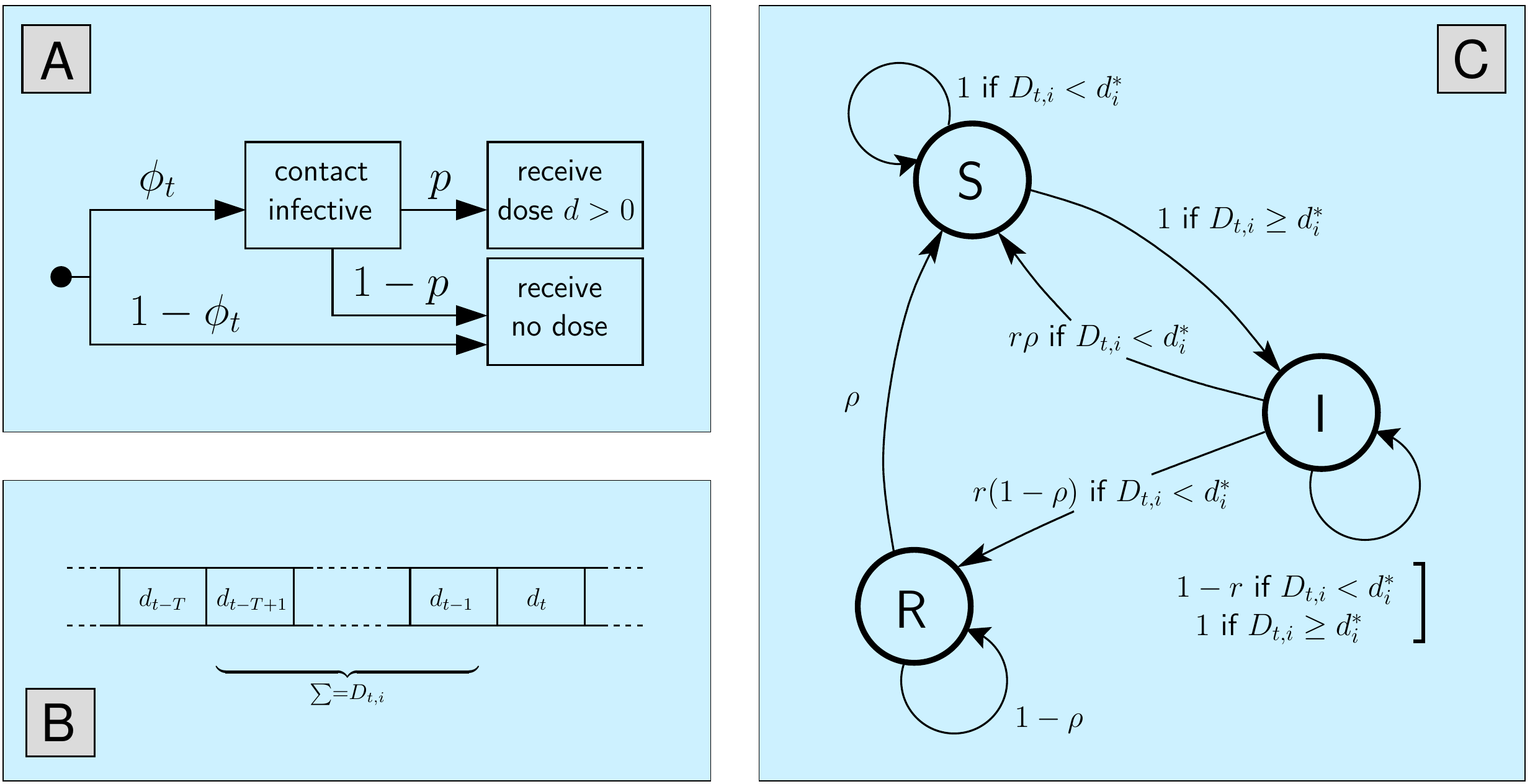}
  \caption{
    (A) 
    Representation of how an individual's dose at time $t$ is determined.
    At each time step, each individual, \textit{regardless} of its state, contacts 
    one other randomly chosen individual.
    The probability that individual $i$ contacts an infective is then
    $\phi_t$, i.e., the current fraction of infectives in the
    population.  If the contact is infected
    then with probability $p$, $i$ is exposed to the contagious entity
    and receives a dose $d$ drawn from a fixed distribution $f$.
    Otherwise and also if the contacted individual is not infected
    in the first place (occurring with probability $1-\phi_t$),
    $i$ receives a dose of zero size.
    (B) 
    Individual $i$ then updates its dose count $D_{t,i}$ by `forgetting' the dose it received $T+1$ time
    steps ago and incorporating the current dose [\Req{gcontlong.eq:sumT}].
    (C)
    Transition probabilities for individual $i$ cycling through
    the three states S (susceptible), I (infected), and R (recovered).
    If $i$ is in the susceptible state, it becomes infected 
    with probability 1 once its dose count $D_{t,i}$
    exceeds its threshold $\dstari$,
    otherwise it remains susceptible.  If $i$ becomes infected,
    then whenever $i$'s dose count
    drops below its threshold, it has a probability
    $r$ of recovering in each time step.  Once $i$ is in state R
    (where it is immune to infection), it becomes susceptible again 
    with probability $\rho$, again in each time step.
    Note that if $r=\rho=1$, infected individuals whose dose
    count falls below their threshold immediately return
    to the susceptible state.
  }
  \label{gcontlong.fig:SIRtransitions}
\end{figure*}

Our model, aspects of which we have reported in brief
elsewhere~\citep{dodds2004pa}, comprises a population of $N$
individuals, each of which is assumed to occupy
one of three states: \textit{susceptible} (S);
\textit{infected} (I); or \textit{removed} (R). 
Figure ~\ref{gcontlong.fig:SIRtransitions} 
provides a schematic representation of our model which
we describe as follows.
At each (discrete) time step $t$,
each individual $i$ comes into contact with another individual chosen
uniformly at random from the population.
If the contact is infected---an event that occurs with probability $\phi_t$, the fraction
of infectives in the population---then with probability $p$, $i$ receives
a `dose' $d$ drawn from a fixed dose-size distribution $f$; else
$i$ receives a dose of size $0$. We call a successful transmission of a
positive dose an \textit{exposure} and $p$ the \textit{exposure probability}. 
Individuals carry a
\textit{memory} of doses received from their last $T$ contacts and we
denote the sum of individual $i$'s last $T$ doses ($i$'s \textit{dose count}) 
at the $t$th time step by
\begin{equation}
  \label{gcontlong.eq:sumT}
  D_{t,i} = \sum_{t'=t-T+1}^{t} d_{t',i}.
\end{equation}
If $i$ is in the susceptible state, then it becomes infected once $D_{t,i}$ exceeds
$i$'s \textit{dose threshold} $\dstari$, where $\dstari$ is drawn from a given
distribution $g$ (dose thresholds do not change with time). Note
that we differentiate exposure from infection, the latter being the
possible result of one or more exposures and only occurring once a
susceptible individual's threshold has been equaled or
exceeded.  Having become infected, an individual remains in state I until its dose
count drops below its threshold, at which point it recovers with
probability $r$ at each time step.  Once recovered, an individual
returns to being susceptible with probability $\RtoSprob$, again at
each time step.  

\begin{figure}[tbp]
  \centering
  \includegraphics[width=0.49\textwidth]{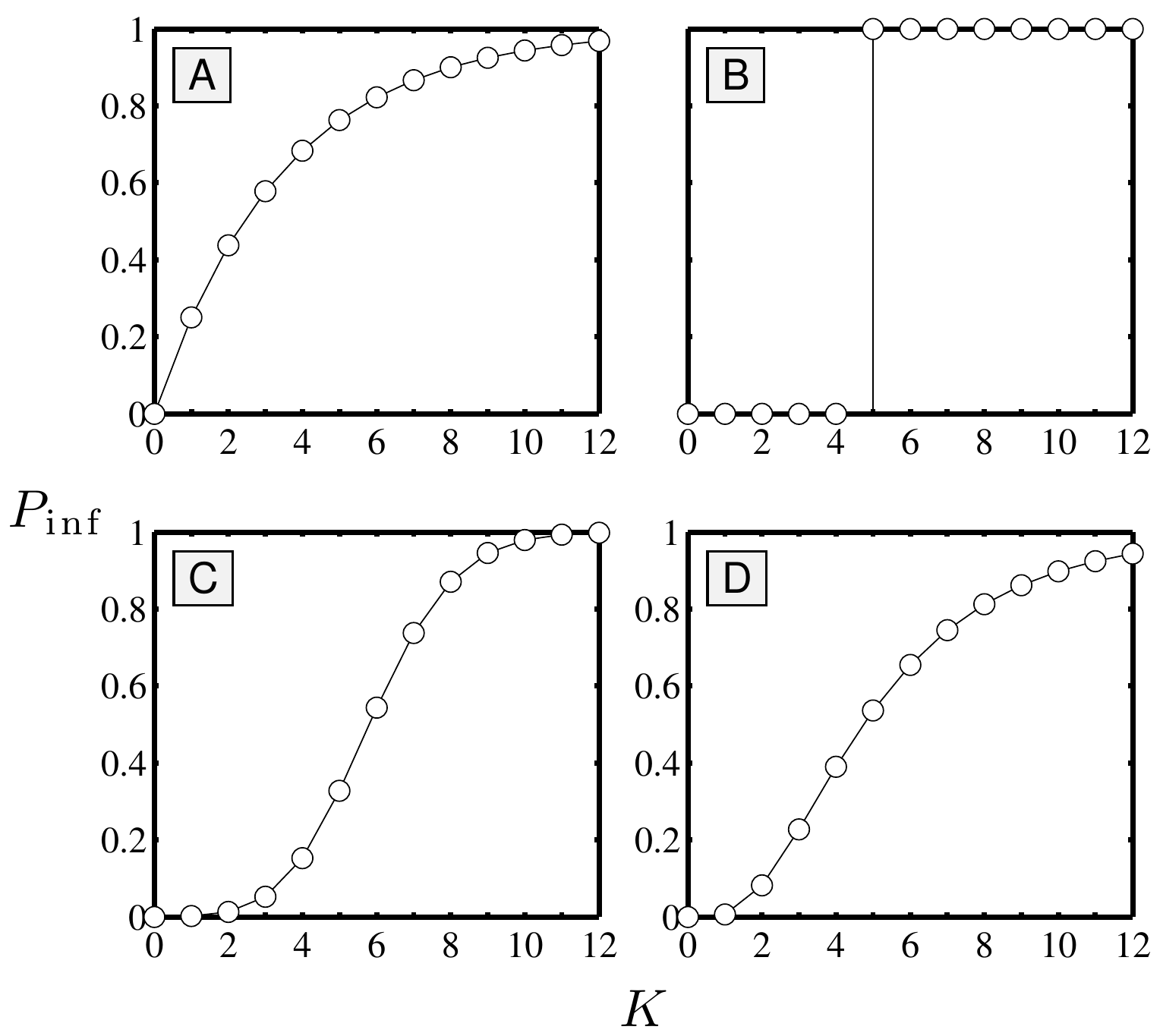}
  \caption{
    Examples of dose response, i.e., 
    $P_{\rm inf}(K)$, the probability an individual becomes infected due to
    $K$ exposures in the last $T (=12)$ time steps 
    [see \Req{gcontlong.eq:Pinfbare}].
    The plots correspond to 
    (A) independent interaction (or disease-like) models with homogeneous
    dose sizes and thresholds 
    [$p=0.25$, dose distribution $f(d) = \delta(d-1)$,
    and threshold distribution $g(\dstar)=\delta(\dstar-1)$];
    (B) deterministic threshold models with homogeneous
    dose sizes and thresholds 
    [$p=1$, $f(d) = \delta(d-1)$, and $g(\dstar)=\delta(\dstar-5)$];
    (C) stochastic threshold models with
    heterogeneous doses and homogeneous thresholds
    [$p=0.9$, doses distributed lognormally with unit
    mean and standard deviation $0.5$, and $g(\dstar)=\delta(\dstar-5)$];
    and
    (D), same as (C) but with heterogeneity in thresholds introduced via
    a lognormal distribution with mean of 5 as per (C) and standard deviation of 10.
  }
  \label{gcontlong.fig:doseresponse30}
\end{figure}

The probability $P_{\rm inf}$ that a susceptible individual 
who comes into contact with $K$ infected individuals in $T$
time steps will become infected
is therefore given by
\begin{equation}
  \label{gcontlong.eq:Pinfbare}
  P_{\rm inf}(K)
  =
  \sum_{k=1}^{K}
    \binom{K}{k}
    p^k
    (1-p)^{K-k}
    P_k,
\end{equation}
where $K=1,\ldots,T$, and 
\begin{equation}
  \label{gcontlong.eq:phifixgendosedstar0}
  P_k
  = 
  \int_{0}^{\infty} \dee{\dstar}
  g(\dstar)
  \int_{d=\dstar}^{\infty} \dee{d}\, \dosedistk(d)
\end{equation}
Both $P_{\rm inf}(K)$ and $P_k$ are important quantities in our model.
The quantity $P_k$ is the expected fraction of a population that will be infected
by $k$ exposures, where the distribution $\dosedistk(d)$, the $k$-fold
convolution of $f$, is the probability that the sum of $k$ doses
will be equal to $d$.  
The infection probability $P_{\rm inf}(K)$ gives, in effect, a
\textit{dose response curve}~\citep{haas2002} averaged over all members
of the population and also the distribution of dose sizes (where we
note that $K$ contacts with infected individuals will result in $k$
actual exposures with probability $\binom{K}{k} p^k (1-p)^{K-k}$).
Figure~\ref{gcontlong.fig:doseresponse30} shows examples of dose
response curves, calculated from \Req{gcontlong.eq:Pinfbare}
for four configurations of the model. The plots correspond to (A)
independent interaction, (B) deterministic threshold, and, in both (C) and (D),
stochastic threshold models.  For the independent interaction example, a single
exposure is needed to generate an infection and so exposures
effectively act independently.  The deterministic threshold example
incorporates uniform dose sizes and thresholds, and 
when the probability of an exposure is set to $p=1$, the
response becomes deterministic (individuals are always infected when
their dose count is met or exceeded and never otherwise); but now
the threshold can only be exceeded by multiple infections.
The two
stochastic cases generalize the deterministic case by allowing (C)
dose sizes to be heterogeneous, and (D) both dose sizes and thresholds
to be heterogeneous.

We explore the behavior of our model with respect to
three qualitative types of dynamics: (1) permanent removal ($\RtoSprob=0$)
dynamics, analogous to so-called SIR models in mathematical
epidemiology in which individuals either die or acquire permanent
immunity; (2) temporary removal ($1>\RtoSprob>0$) dynamics, analogous to
SIRS models where recovered individuals become susceptible again after
a certain period of immunity; and (3) instantaneous replacement ($\RtoSprob=1$)
dynamics, analogous to SIS models, where infected individuals
immediately become susceptible again upon recovery. Chicken pox, for
example, would correspond to SIR-type contagion, while the common cold
resembles the SIRS case.  Because of its simplicity,
we obtain the majority of our analytic results for the somewhat special SIS case
($\RtoSprob=1$).  However, our main findings for the SIS case have analogs in
the more complicated SIRS and SIR cases which we investigate with numerical
simulations.  Furthermore, while the
assumption of instantaneous re-susceptibility is probably not
appropriate in the case of contagious biological agents (where
recovery is generally associated with some period of immunity), it may
well be approximately true for contagious social influences, such as
``social smoking,'' where an individual, having quit, can restart
immediately. A summary of the main parameters of the model and their
definitions is provided in Table~\ref{gcontlong.tab:defns}.

\begin{table}[htbp]
  \begin{center}
    \begin{tabular}{c|l}
      \hline
      $T$ & length of memory window \\
      $p$ & probability of exposure given contact with infective\\
      $r$ & probability of moving from infected to recovered state\\ %, $P(I \rightarrow R)$\\
      $\RtoSprob$ & probability of moving from immune to susceptible state \\ %, $P(R \rightarrow S)$\\
      $f(d)$ & distribution of dose sizes $d$ \\
      $g(\dstar)$ & distribution of individual thresholds $\dstar$\\
      $\bar{\dstar}$ & uniform threshold of homogeneous population \\
      $\phi_t$ & fraction of population infected at time $t$ \\
      $\phifix$ & steady-state fraction of population infected \\
      \hline
    \end{tabular}
    \caption{
      Summary of main model parameters and definitions.
      }
    \label{gcontlong.tab:defns}
  \end{center}
\end{table}

We structure the remainder of the paper as follows. In
section~\ref{gcontlong.sec:homog}, through analysis and simulations,
we examine in detail the SIS $(\RtoSprob=1)$ version of the model for a
homogeneous population, where by ``homogeneous,'' we mean that all doses
are equal and of unit size [i.e., $f(d)=\delta(d-1)$], and that all individuals
have the same threshold [i.e., $g(\dstar) =
\delta(\dstar-\bar{\dstar})$]. For homogeneous populations, we find
that only two universal classes of dynamics are possible: (1)
\textit{epidemic threshold dynamics}~\footnote{
  In this paper, we use the word `threshold' in two terms:
  `threshold models' and `epidemic threshold models.'
  At the risk of some confusion, we have done so to maintain consistency
  with the two distinct literatures we are connecting, sociology and mathematical epidemiology.
  Threshold models of sociology refer to individual level thresholds whereas
  the term epidemic threshold refers to the critical reproduction number of a disease,
  above which an outbreak is assured.
},
according to which initial
outbreaks either die out or else infect a finite fraction of the
population, depending on whether or not the infectiousness $p$ exceeds
a specific critical value $p_c$; 
and (2) \textit{critical mass dynamics}
according to which a finite fraction of the population can only ever
be infected in equilibrium if the initial outbreak size itself
constitutes a finite ``critical mass.''  Although homogeneity is
a restrictive assumption for biological or social
contagion, it provides a useful special case in that it illuminates
the basic intuitions required to understand the more general,
heterogeneous case.  In section~\ref{gcontlong.sec:heterog}, we relax
the homogeneity assumption and move to the richer and more realistic
case of arbitrarily distributed dose sizes and individual thresholds.
Here we find that three universal classes of contagion models are
possible.  In addition to the epidemic threshold and critical
mass classes that carry over from the homogeneous case, we also find an
intermediate class of \textit{vanishing critical mass dynamics} in
which the size of the required critical mass diminishes to zero for
$p<1$. Furthermore, we determine where the transitions between these
classes occur and also the conditions required for more complicated
kinds of contagion models to arise. Subsequently, in
section~\ref{gcontlong.sec:SIR}, we explore the SIRS and SIR versions
of the model, finding behavior that in many ways resembles that of
the simpler SIS case.  In the SIR case, for example, it is necessarily
true that all epidemics eventually burn themselves out (because for
$\RtoSprob=0$, the removed condition is an absorbing state).  However, we
find that the presence of memory may cause an epidemic to persist for a
surprisingly long time.
In section~\ref{gcontlong.sec:concl}, we conclude our
analysis, discussing briefly the implications of our findings for
stimulating or retarding different contagious processes.  Finally, in
the appendices we provide detailed derivations of the analytical
results from sections~\ref{gcontlong.sec:homog}
and~\ref{gcontlong.sec:heterog}.
%, with 
%Appendix~\ref{gcontlong.app:exactdstar=2} being of some mathematical interest.

\section{Homogeneous SIS contagion models}
\label{gcontlong.sec:homog}

\subsection{Epidemic threshold models}
\label{gcontlong.subsec:epithreshmodels}

We begin our analysis with a simple non-trivial case
of our model, for which we assume that individuals are:
identical (i.e., the population is homogeneous);
have no memory of past interactions $(T=1)$; 
and, upon recovery from infection, immediately
return to the susceptible state (i.e., $\rho=1$). 
In this limit our model coincides with the SIS version of the
traditional Kermack-McKendrick model~\citep{kermack1927,murray2002}, as
individuals with no memory necessarily become infected upon exposure
to a single infected individual [$g(\dstar)=\delta(\dstar-1)$].  For a
specified recovery time $(r\leq 1)$ (which again is identical across
all individuals), the fraction of infected individuals at time $t$,
$\phi_{t}$, evolves according to
\begin{equation}
  \label{gcontlong.eq:r<1k1T1pre}
  \phi_{t+1} = p \phi_{t}
  + \phi_t(1 - p\phi_t)(1-r).
\end{equation}
The first term on the right is the fraction of individuals
newly infected between time $t$ and $t+1$, regardless of
whether or not they were infected beforehand (the model
allows for individuals to recover and be reinfected
within one time step).  The second term 
is the fraction of individuals who were infected in the 
preceding time step, were not infected between time $t$ and $t+1$, and did not recover.

\begin{figure}[tbp!]
  \centering
  \includegraphics[width=0.49\textwidth]{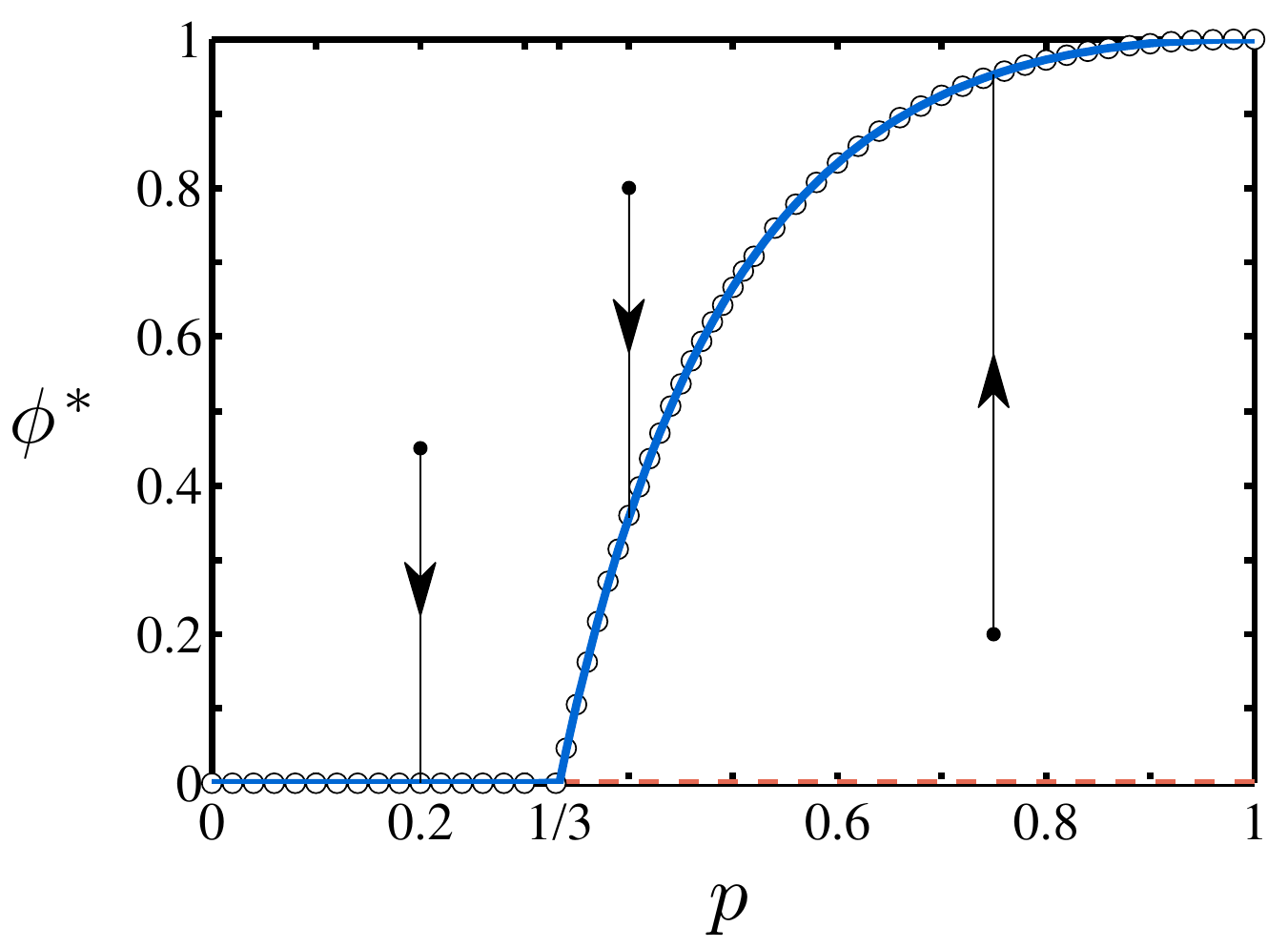}                                        
  \caption{
    Fixed point curves for an example epidemic threshold model.
    The stable fixed point curves shown are from 
    simulation (circles) and \Req{gcontlong.eq:r<1k1T>1}
    (blue line) for $\bar{\dstar}=1$, $T=2$, and $\tau=1/r-1=1$.
    Unstable fixed points are indicated by the dashed red line.
    The trajectories of three initial conditions are shown
    to illustrate how the level of an epidemic evolves for different values of $p$.
    The epidemic threshold $p_c=1/3$ is as predicted by 
    $p_c = (T + \tau)^{-1}$, \Req{gcontlong.eq:r<1k1T>1pc}.
    For the simulation, the population size $N=10^5$, the
    number of time steps $N_t=10^3$, each data point
    represents the average value of $\phifix(p)$ over the last $N_s=100$
    time steps, and the initial condition is that all are infected.
  }
  \label{gcontlong.fig:gc_rvar_k1_Tvar}
\end{figure}

Setting $\phi_t=\phifix$ in \Req{gcontlong.eq:r<1k1T1pre},
we find the stable fixed points of the model as a function of $p$ are given by
\begin{equation}
  \label{gcontlong.eq:r<1k1T1}
\phifix = 
\left\{
  \begin{array}{ll}
    0 & 
    \mbox{for} \quad 0 \le p \le r, \\
    \frac{1 - r/p}{1-r} & 
    \mbox{for} \quad r < p \le 1.
  \end{array}
\right.
\end{equation}
Furthermore, a single unstable set of fixed points is found along
$\phifix=0$ for $r < p \le 1$.  Thus, the standard memoryless SIS
model exhibits an \textit{epidemic threshold}~\citep{murray2002} at
$p=p_c$, as displayed graphically in
Figure~\ref{gcontlong.fig:gc_rvar_k1_Tvar}: for $p \le p_c = r$ no
infection survives ($\phifix \equiv 0$) while for $p > p_c = r$, a
stable, finite fraction of the population will become
infected ($\phifix > 0$). In the language of dynamical systems theory,
the epidemic threshold is a transcritical bifurcation~\citep{strogatz1994}
which lies at the intersection of two fixed point curves,
where the stable curve changes to unstable at
the intersection and vice versa (in statistical mechanics,
such behavior is referred to as a second-order or continuous
\textit{phase transition}, where $(p_c,0)$ is called the
\textit{critical point}~\citep{goldenfeld92}).
For the choice of parameters above, one
branch of the transcritical bifurcation is the $p$-axis which is stable
to the left of $p_c$ and unstable to the right (imagining the
unphysical extension of $\phifix$ to $\phifix<0$, the other branch may
be seen to extend from below the $p$-axis where it is unstable to
above the $p$-axis where it is stable).

We observe this kind of bifurcation structure (i.e., where the rising branch has
positive slope and comprises stable equilibria) to be 
a robust feature with respect to a range of parameter choices, and
we classify all such models as \textit{epidemic threshold models}.
As we show below, the same qualitative equilibrium behavior is present in 
all homogeneous models with dose thresholds set at $\dstar=1$,
even with arbitrary memory $T \ge 1$ and recovery rate $r \le 1$. 
While some details of the dynamics do change as $T$ and $r$ are varied,
the existence of a single transcritical bifurcation depends only on the
assumption that all individuals exhibit ``trivial
thresholds:'' $g(\dstar)=\delta(\dstar-1)$.  

First, allowing memory to be arbitrary ($T>1$) but keeping $r=1$,
we observe that the individuals who are infected at some time $t$ are
necessarily those who have experienced at least one infectious event
in the preceding $T$ time steps; thus we obtain the following implicit
equation for $\phifix$,
\begin{equation}
  \label{gcontlong.eq:r1k1T>1}
  \phifix = 1 - (1 - p\phifix)^T.
\end{equation}
As with the $T=1$ case above, the equilibrium behavior exhibits a
continuous phase transition and therefore also falls into the epidemic
threshold class. While we can no longer find a general, closed-form
solution for $\phifix$, \Req{gcontlong.eq:r1k1T>1} can be rearranged
to give $p$ as an explicit function of $\phifix$:
\begin{equation}
  \label{gcontlong.eq:r1k1T>1_inv}
  p = \phifix^{-1} [ 1 - (1-\phifix)^{1/T} ].
\end{equation}
Taking the limit of $\phifix \rightarrow 0$ in either
\Req{gcontlong.eq:r1k1T>1} or \Req{gcontlong.eq:r1k1T>1_inv}, we find
$p_c=1/T$. 
Thus, receiving at
least one exposure from the last $T$ contacts is analogous to the
zero memory ($T=1$), variable $r$ case above where recovery occurs on a time scale $1/r
\simeq T$.

Next, we also allow the recovery rate to be arbitrary $(r<1)$. 
To find the fixed point curves, we modify
\Req{gcontlong.eq:r1k1T>1} to account for the fraction of individuals
that have not experienced a single exposure for at least the last $T$
time steps but have not recovered from a previous infection.  We first
write down the probability that an individual last experienced an
infectious event $m$ time steps ago and has not yet recovered.
Denoting the sequence of a positive unit dose followed by $m$ 0's as
$H_{m+1}$, we have
\begin{equation}
  \label{gcontlong.eq:k=1_b}
  P(\mbox{infected}|H_{m+1}) = p\phifix (1 -p\phifix)^{m}
  (1-r)^{m-T+1},
\end{equation}
The first term on the right hand side of the \Req{gcontlong.eq:k=1_b}
is the probability of a successful exposure; the second term is the
probability of experiencing no successful exposures in the subsequent
$m$ time steps; and the final term is the probability that, once the
memory of the initial single exposure has been lost after $T$ time
steps, the individual remains infected.  Since we are only concerned
with individuals who have `forgotten' the source of the infection, we
have $m \ge T$.  Summing over $m$, we obtain the total probability
that an individual was infected at least $T$ time steps ago and has
not yet recovered:
\begin{equation}
  \label{gcontlong.eq:k=1inf}
  \sum_{m=T}^{\infty} P(\mbox{infected}|H_{m+1}) = \frac{p\phifix
  (1-p\phifix)^T (1-r)}{1 - (1-p\phifix)(1-r)}
\end{equation}
Adding this fraction to the right hand side of
\Req{gcontlong.eq:r1k1T>1} then gives
\begin{eqnarray}
  \label{gcontlong.eq:r<1k1T>1}
  \phifix & = & 1 - (1-p\phifix)^T \left[
    1 -
    \frac{p\phifix(1-r)}
    {1 - (1-p\phifix)(1-r)}
  \right], \nonumber \\
  & = & 1 - \frac{r (1-p\phifix)^T }
    {1 - (1-p\phifix)(1-r)}.
\end{eqnarray}
Figure~\ref{gcontlong.fig:gc_rvar_k1_Tvar} shows a comparison
between the above equation and simulation results
for $T=2$, $r=1/2$, and $\bar{\dstar}=1$.

Taking the limit of small $\phifix$ in \Req{gcontlong.eq:r<1k1T>1}
we find the epidemic threshold to be
\begin{equation}
  \label{gcontlong.eq:r<1k1T>1pc}
  p_c = \frac{r}
    {1 + r(T-1)} = \frac{1}{T + 1/r - 1}.
\end{equation}
Checking the special cases of the preceding calculations,
we find $p_c=r$ when $T=1$ and $p_c=1/T$ when $r=1$.
Denoting the mean time to recovery
of an infected, isolated individual by $\tau$,
we observe $\tau=1/r-1$ and \Req{gcontlong.eq:r<1k1T>1pc} becomes
\begin{equation}
  \label{gcontlong.eq:r<1k1T>1pc2}
  p_c = \frac{1}{T + \tau}.
\end{equation}
The time scales $T$ and $\tau$ can thus both be thought of as 
corresponding to two different kinds of memory, 
the sum of which---the total characteristic time scale of memory in the
model---determines the position $p_c$ of the epidemic threshold.
Qualitatively, therefore, all homogeneous models that possess trivial individual
thresholds exhibit the same kind of equilibrium dynamics.  Varying the
thresholds, however, produces equilibrium behavior of a quite distinct
nature, as we show in the next section.

\subsection{Critical mass models}
\label{gcontlong.subsec:critmassmodels}

\begin{figure}[tbp!]
  \centering
  \includegraphics[width=0.49\textwidth]{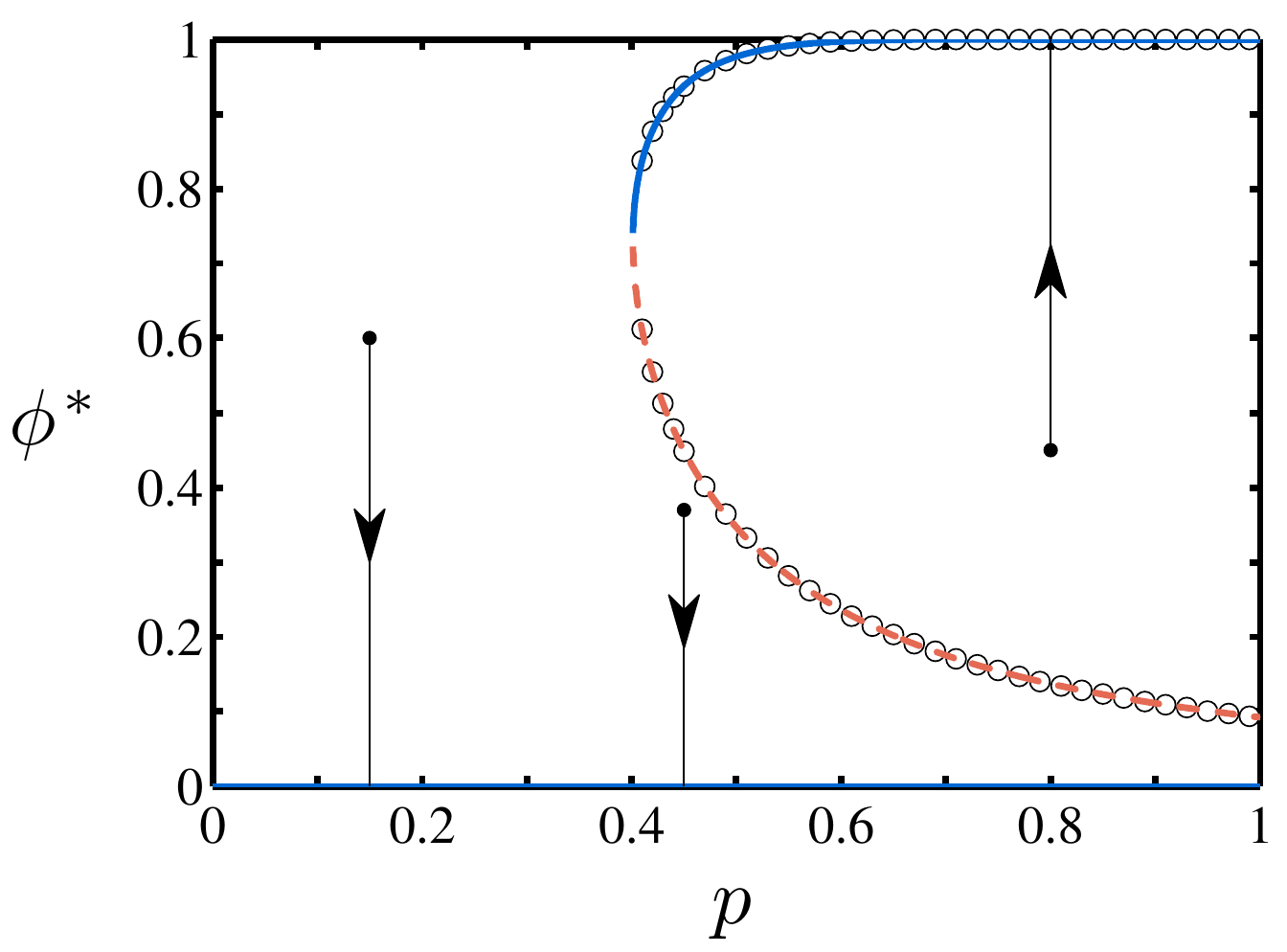}
  \caption{
    Comparison between simulation (circles) and theory (lines, \Req{gcontlong.eq:r=1T>1k>1}) for
    $T=12$, $\bar{\dstar}=3$, and $r=1$. 
    For $\bar{\dstar}>1$, the homogeneous SIS contagion model
    exhibits a saddle-node bifurcation.
    Shown are the non-zero stable and unstable points.
    For the theoretical curves, solid lines represent stable points
    and dashed ones unstable points.
    All points on the line $\phifix=0$ are stable points.
    The arrows show trajectories of the system for
    three example initial conditions.
    Simulations details are as per Figure~\ref{gcontlong.fig:gc_rvar_k1_Tvar}.
    The location of the unstable fixed point curve is determined by binary search.
  }
  \label{gcontlong.fig:gc_r1_k3_T12}
\end{figure}

When individuals of a homogeneous population require more than one
exposure to become infected---that is, 
when $g(\dstar) = \delta(\dstar-\bar{\dstar})$ and $\bar{\dstar}>1$---the closed-form
expression for the fixed points $\phifix$ in the case of $r=1$ is
\begin{equation}
  \label{gcontlong.eq:r=1T>1k>1}
  \phifix = 
  \sum_{i=\bar{\dstar}}^{T}
  \binom{T}{i}
  (p\phifix)^{i} (1 - p\phifix)^{T-i}.
\end{equation}
We now observe a fundamentally different behavior of the model: the
transcritical bifurcation that is characteristic of epidemic threshold
models is absent and is replaced by a saddle-node
bifurcation~\citep{strogatz1994}, an example of which is illustrated in
Figure~\ref{gcontlong.fig:gc_r1_k3_T12}.  We call models whose
equilibrium states are determined in this manner \textit{critical mass
models} because, as indicated by the arrows in
Figure~\ref{gcontlong.fig:gc_r1_k3_T12}, an epidemic will not spread
from an infinitesimal initial outbreak, requiring instead a finite
fraction of the population, or ``critical mass''~\citep{schelling1978},
to be infected initially.  Saddle-node bifurcations (or backwards bifurcations) 
have also been observed in a number of unrelated epidemiological 
multi-group models, arising from differences between groups and inter-group 
contact rates~\cite{hadeler1995,hadeler1997,dushoff1998,kribs-zaleta2000,greenhalgh2000,kribs-zaleta2002}.

\begin{figure}[tbp!]
  \centering
    \includegraphics[width=0.49\textwidth]{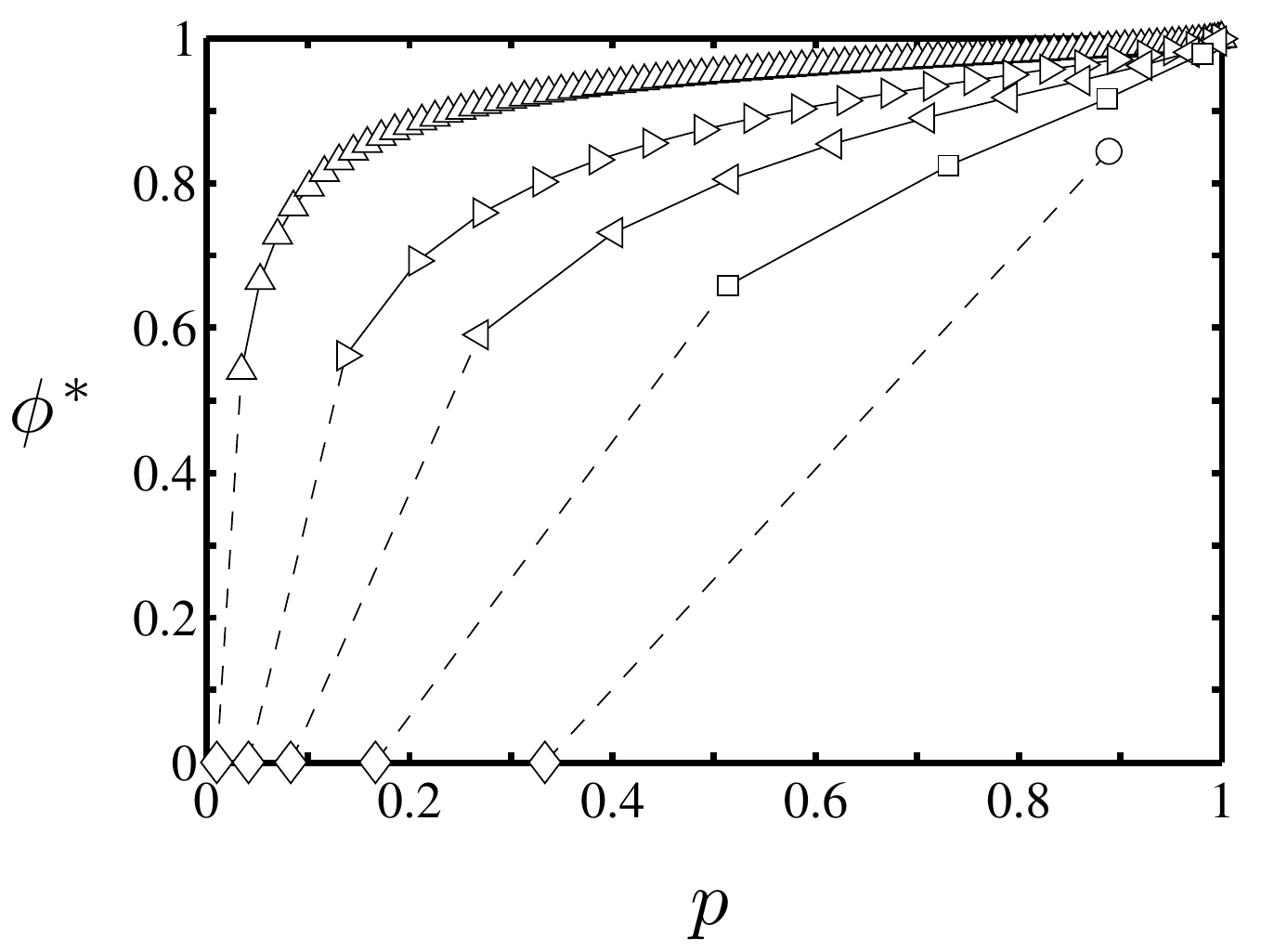}
    \caption{
      Bifurcation points for $\bar{\dstar}>1$, $\phifix>1$, and $r=1$,
      determined numerically using equations~\Req{gcontlong.eq:r=1T>1k>1}
      and~\Req{gcontlong.eq:r=1T>1k>1diff2}.  The data shown are for
      $T=3$ ($\bigcirc$), 
      $T=6$ ($\Box$),
      $T=12$ ($\triangleleft$),
      $T=24$ ($\triangleright$),
      and
      $T=96$ ($\vartriangle$).
      The solid lines guide the eye for the range $\bar{\dstar}=2$ to $\bar{\dstar}=T-1$
      with $\bar{\dstar}$ increasing with $p_b$.
      The dashed lines connect to the transcritical bifurcation points 
      ($\diamond$) observed for $\bar{\dstar}=1$.  
      Note that no bifurcations occur when $\bar{\dstar}=T>1$ (the sole fixed point is $\phifix=1$ when $p=1$)
      and when $\bar{\dstar}=T=1$ (the fixed points lie along the line $p=1$ and $0 \le \phi \le 1$).
      }
    \label{gcontlong.fig:gc_bipts_r1}
\end{figure}

Although it is not generally possible to obtain an expression for $\phifix(p)$
from \Req{gcontlong.eq:r=1T>1k>1}, we are able to write down
a closed-form expression involving the position of the 
saddle-node bifurcation $(p_b,\phifixb)$ 
(see Appendix~\ref{gcontlong.app:saddlenodeextra} for details):
\begin{equation}
  \label{gcontlong.eq:r=1T>1k>1diff2}
  0 = 
  \sum_{i=\bar{\dstar}}^{T} 
  \binom{T}{i}
  z^{i-\bar{\dstar}} (1 - z)^{T-i-1}
  [i -1 - z(T-1)],
\end{equation}
where $z = p_b \phifixb$.
Using~\Req{gcontlong.eq:r=1T>1k>1diff2}, we solve for
$z$ by standard numerical means and then use~\Req{gcontlong.eq:r=1T>1k>1}
to obtain $\phifixb$ and hence $p_b$.
Figure~\ref{gcontlong.fig:gc_bipts_r1} shows positions of saddle-node
bifurcation points computed for a range of values of $T$ and
$\bar{\dstar}$. In all cases, as $\bar{\dstar}$ increases, the
bifurcation point moves upward and to the right of the fixed point
diagram.  However, for small values of $\bar{\dstar}$ and $T$,
we are able to determine $(p_b,\phifixb)$ exactly.
For example, for $\bar{\dstar}=2$ and $T=3$, we find that $p_b=8/9$ and
$\phifixb=27/32$ and that the bifurcation is parabolic.  

\begin{figure}[tbp!]
  \centering
  \includegraphics[width=0.49\textwidth]{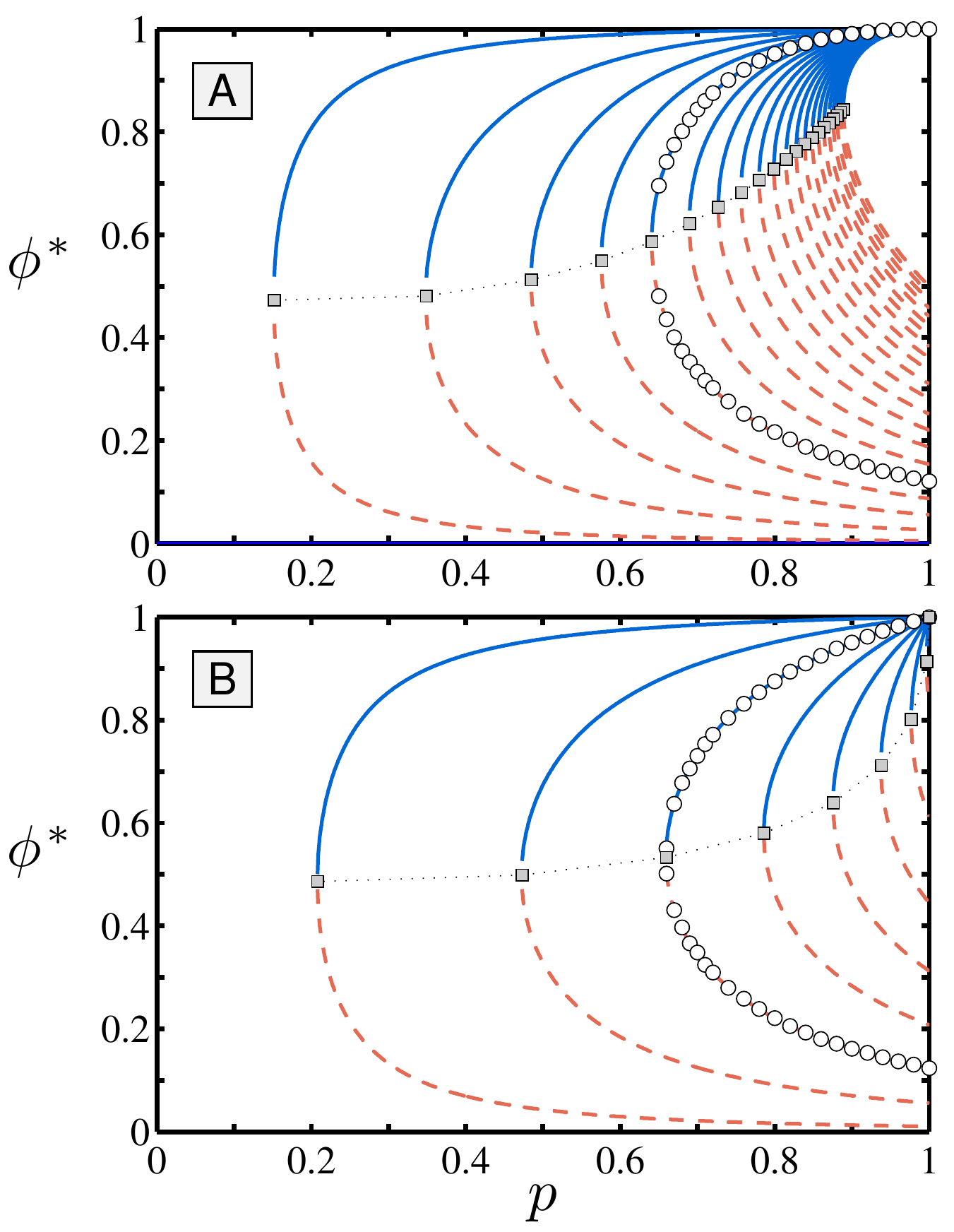}
  \caption{
    A: Theory versus simulation for $T=3$, $\bar{\dstar}=2$, and varying $r$.  
    Solid lines represent theoretically derived stable fixed points,
    dashed lines represent unstable ones, and the squares indicate bifurcation points
    (using Eqs.~\req{gcontlong.eq:Gamma2} and~\req{gcontlong.eq:Gammafull}).
    From left to right, we have $r=0.01, 0.05, 0.10, 0.15, \ldots, 1.00$.
    The circles correspond to simulation data for the case $r=0.2$
    and matches the theoretical curves.  
    B: Theoretical fixed point curves for $T=2$, $\bar{\dstar}=2$, and varying $r$.  
    Bifurcations appear for $r < 0.3820 \pm 0.0001$.  Otherwise, $(p,\phifix)=(1,1)$ is
    the only fixed point of the system with $\phifix>0$.  The values of $r$ are as per the $T=3$ case in A.
    Circles correspond to simulation results for $r=0.1$.
    For all values of $r$, both of the above systems possess a line of stable 
    fixed points described by $0 \le p < 1$ and $\phifix=0$, i.e., the $p$-axis.
  }
  \label{gcontlong.fig:gc_T3_k2_bif_theorycomp2}
\end{figure}

As with epidemic threshold models, the analysis for critical mass
dynamics can be generalized to include a non-trivial recovery rate
$r<1$.  Now, however, we are unable to find a closed-form
expression for the fixed points for general $T$ and
$\bar{\dstar}$.
The difficulty in making such a computation lies in finding
an expression for the number of individuals whose dose counts
are below threshold but are still infected since they have not yet recovered.
For the $\bar{\dstar}=1$ case, this was straightforward since 
the only way to stay below threshold
was to experience a sequence of null exposures. 
Nevertheless, for $r<1$, we can formally modify the expression for $\phifix$ 
given in \Req{gcontlong.eq:r=1T>1k>1}:
\begin{equation}
  \label{gcontlong.eq:Gamma2}
  \phifix = \Gamma(p,\phifix;r,T) + \sum_{i=\bar{\dstar}}^{T}
  \binom{T}{i} (p\phifix)^{i} (1 - p\phifix)^{T-i},
\end{equation}
where the additional term $\Gamma(p,\phifix;r)$ accounts for the
proportion of below threshold individuals who have not yet recovered.
For small values of $T$ and $\bar{\dstar}$, exact expressions for
$\Gamma(p,\phifix;r)$ can be derived and then $\phifix$ can be solved
for numerically. In Appendix~\ref{gcontlong.app:exactdstar=2}, we
consider two special cases, $\bar{\dstar}=2$ for $T=2$ and
$T=3$, that illustrate the process of constructing expressions for
$\Gamma(p,\phifix;r)$. 
These results allow us to explore the movement of the fixed points with decreasing $r$.
Figures~\ref{gcontlong.fig:gc_T3_k2_bif_theorycomp2} A and B show
comparisons over a range of $r$ between the solutions of
\Req{gcontlong.eq:Gamma2} and simulations for $T=2$ and
$T=3$ respectively, confirming that the agreement is excellent.

\section{Heterogeneous SIS contagion models}
\label{gcontlong.sec:heterog}

\subsection{Distributions of doses and thresholds}
\label{gcontlong.subsec:dosethreshdist}

In real populations, both in the context of
infectious diseases and also social influences, individuals
evidently exhibit varying levels of susceptibility.  
Furthermore, contacts between infected and susceptible
individuals can result in effective exposures of variable size,
depending on the individuals in question, as well as the nature of 
their relationship and the circumstances of the contact (duration,
proximity, etc.).  Our homogeneity assumption of the preceding section
is therefore unlikely to be justified in any real application.  
Furthermore, as we show below, 
the more general case of heterogeneous populations, while preserving much of the
structure of the homogeneous case, also yields a new class of dynamical
behavior; thus the inclusion of heterogeneity not only makes the model
more realistic, but also provides additional qualitative insight.

While in principle all parameters in the model could be assumed to
vary across the population, we focus here on two
parameters---individual threshold $\dstar$ and dose size $d$---which
when generalized to stochastic variables, embody the variations in
individual susceptibility and contacts that we wish to
capture. We implement these two sources of heterogeneity
as follows.  In the case of thresholds, each individual
is assigned a threshold drawn randomly from a specified probability
distribution $g(\dstar)$ at $t=0$ which remains fixed for all $t$.
In effect, this assumption implies that individual characteristics remain roughly invariant on the time
scale of the dynamics, rather than varying from moment
to moment.  By contrast, in order to capture the unpredictability of circumstance,
we assign dose sizes stochastically,
according to the distribution $f(d)$, independent both of time
and the particular individuals between whom the contact occurs.

Again considering first the more tractable $r=1$ version of the SIS
model, we first examine the effect of allowing dose size $d$ to vary while
holding thresholds $\dstar$ fixed across the population.  In the
homogeneous case, $k$ exposures of a susceptible to infected
individuals resulted in a dose count of $k$, but the result can
be more complicated when $d$ is allowed to vary continuously. Note that $\dstar$ also no
longer need be an integer. First, we calculate the probability that a
threshold will be exceeded by $k$ doses. As the distribution of dose
size is now some arbitrary function $f$, we have that the probability
distribution of the sum of $k$ doses is given by the $k$-fold
convolution
\begin{equation}
  \label{gcontlong.eq:kfoldconv}
  \Prob\left(
    \textstyle{\sum_{j=1}^{k} d_j = d}
    \right)
    =
    f \ast f \ast \cdots \ast f (d) = \dosedistk(d).
\end{equation}
The probability of exceeding $\dstar$ is then
\begin{equation}
  \label{gcontlong.eq:kfoldconvexceed}
  \Prob\left(
    \textstyle{\sum_{j=1}^{k} d_j \ge \dstar}
    \right)
    =
    \int_{d=\dstar}^{\infty} \dee{d} \, \dosedistk(d).
\end{equation}
Since, for $r=1$, individuals are only infected when their dose
count exceeds their threshold, we have that the steady-state 
fraction infected is given by
\begin{equation}
  \label{gcontlong.eq:phifixgendose}
  \phifix =
  \sum_{k=1}^{T}
  \binom{T}{k}
  (p\phifix)^{k}
  (1-p\phifix)^{T-k}
    \int_{d=\dstar}^{\infty} \dee{d} \, \dosedistk,
%  \Prob\left(
%    \textstyle{\sum_{i=1}^{k} d_i \ge \dstar}
%    \right),
\end{equation}
where we have averaged over all possible ways
an individual may experience $1 \le k \le T$ exposures in $T$
interactions.

Next, in order to account for any variation $\dstar$,
we must incorporate another layer of 
averaging into~\Req{gcontlong.eq:phifixgendose} as follows:
\begin{align}
  \label{gcontlong.eq:phifixgendosedstar}
  & \phifix =  \nonumber  \\
  & \int_{0}^{\infty} \! \! \! \dee{\dstar}
  g(\dstar) 
  \sum_{k=1}^{T}
  \binom{T}{k}
  (p\phifix)^{k}
  (1-p\phifix)^{T-k} 
 \int_{d=\dstar}^{\infty} \! \! \! \dee{d}\, \dosedistk(d), \nonumber  \\
  & = 
  \sum_{k=1}^{T}
  \binom{T}{k}
  (p\phifix)^{k}
  (1-p\phifix)^{T-k} 
  P_k,
\end{align}
where $P_k$ is defined by \Req{gcontlong.eq:phifixgendosedstar0}.
An important insight that can be derived immediately from 
\Req{gcontlong.eq:phifixgendosedstar} is that all information
concerning the distributions of dose sizes and thresholds is
expressed via the $\{P_k\}$; hence the details of
the functions $f$ and $g$ are largely unimportant.  
In other words, many pairs of $f$ and $g$
can be constructed to give rise to the same $\{P_k\}$ and hence the same
fixed points.  For example, any desired $\{P_k\}$ can be 
realized by a uniform distribution of unit doses $f(d) = \delta(d-1)$,
along with a discrete distribution of thresholds
\begin{equation}
  \label{gcontlong.eq:univPk}
  g(\dstar) = P_1 \delta(\dstar-1) + \sum_{k=1}^{T-1}(P_{k+1}-P_{k}) \delta(\dstar-k-1).
\end{equation}

We observe that for the homogeneous case where doses and thresholds
are fixed at $1$ and $\bar{\dstar}$ respectively [i.e., $f(d) =
\delta(d-1)$ and $g(\dstar) = \delta(\dstar-\bar{\dstar})$], we have
$\dosedistk(d) = \delta(d-k)$ and the expression for $P_k$
[\Req{gcontlong.eq:phifixgendosedstar0}] simplifies to $P_k = 0$ if $k
< \bar{\dstar}$ and $P_k = 1$ if $k \ge \bar{\dstar}$. Substituting
these conditions into \Req{gcontlong.eq:phifixgendosedstar}, we
recover our previous expression \Req{gcontlong.eq:r=1T>1k>1} for the
$r=1$ homogeneous case.

\subsection{Universal classes of contagion}
\label{gcontlong.subsec:universal}
 
\begin{figure*}[tbp!]
  \centering
  \includegraphics[width=0.98\textwidth]{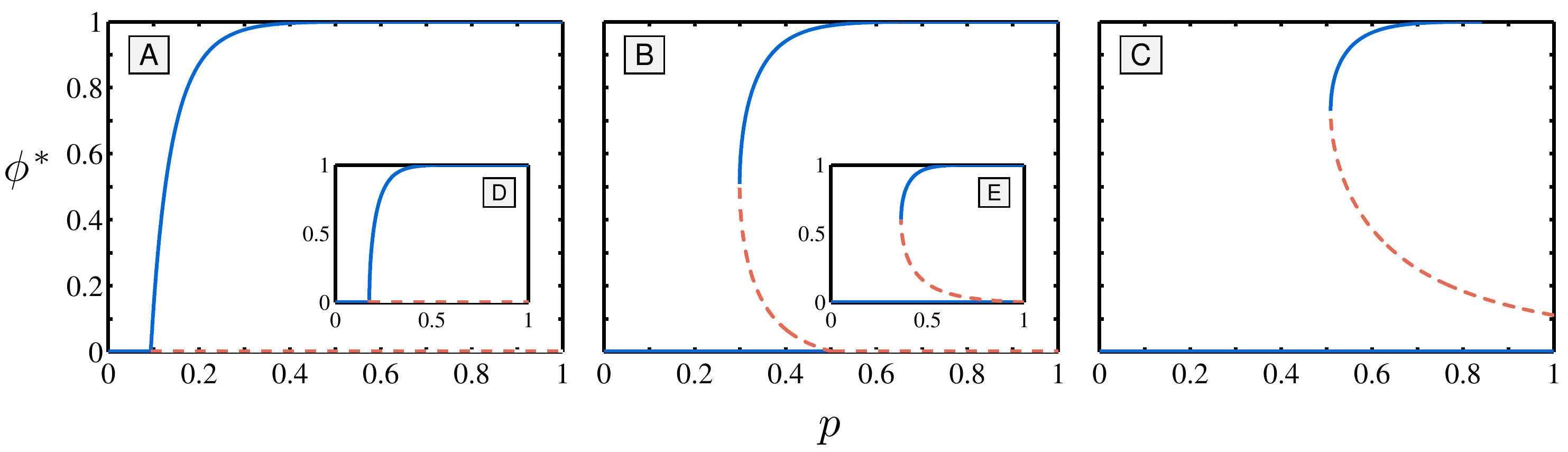}
  \caption{
    Fixed point curves for the three main classes of contagion models produced
    by the model, along with transitions between classes.
    All curves are determined by numerical solution
    of~\Req{gcontlong.eq:phifixgendosedstar} with an error tolerance
    of $10^{-10}$.
    Here, $T=12$, individual doses are lognormally distributed with mean 1
    and a variance of $0.6$ for the underlying normal distribution,
    and thresholds are homogeneously distributed with
    $g(\dstar) = \delta(\dstar-\bar{\dstar})$.
    The main plots correspond to 
    (A) $\bar{\dstar}=0.4$ (class I),
    (B) $\bar{\dstar}=1.5$ (class II),
    and
    (C) $\bar{\dstar}=3$ (class III).
    The insets plots (D) and (E) respectively show fixed point curves for models
    at the class I-class II transition (i.e., when $P_1=P_2/2$ which occurs
    when $\bar{\dstar}=0.8600\ldots$) and the
    class II-class III transition (i.e., when $p_c=1$ which occurs when 
    $\bar{\dstar}=1.9100\ldots$).
    In all plots except (E), the intersection between the fixed point curve
    and the $p$-axis is a transcritical bifurcation.  In plots
    (B), (C), and (E), the second bifurcation is a saddle-node bifurcation.
    }
  \label{gcontlong.fig:threeclasses}
\end{figure*}

In the homogeneous version of the model, we determined the existence
of two classes of dynamics---epidemic threshold and critical
mass---with the former arising whenever $\bar{\dstar}=1$, and the latter
when $\bar{\dstar} \ge 2$.  In other words, in homogeneous populations,
the condition for differentiating between one class of behavior and
another is a discrete one.  Once heterogeneity is introduced, however, we
observe a smooth transition between epidemic threshold and critical
mass models governed by a continuous adjustment of the distributions
$f$ and $d$ (or equivalently the $\{P_k\}$).  One consequence of this now-continuous transition is the
appearance of an intermediate class of models, which we call
\textit{vanishing critical mass} (see
Fig.~\ref{gcontlong.fig:threeclasses}). As with pure critical mass
models, this new class is characterized by a saddle-node bifurcation,
but now the lower unstable branch of fixed points crosses the
$p$-axis at $p_c$; in other words, the required critical mass
``vanishes.''  The collision of the unstable branch of the saddle node
bifurcation with the horizontal axis also effectively reintroduces a
transcritical bifurcation, in the manner of epidemic threshold models.
However, the transcritical bifurcation is different from the one
observed in epidemic threshold models because the rising branch of the
bifurcation has negative, rather than positive, slope and comprises
unstable, rather than stable, fixed points.  Vanishing critical mass
dynamics are therefore qualitatively distinct from both previously
identified classes of behavior, exhibiting important properties of
each: for $p<p_c$, they behave like critical mass models; and for
$p>p_c$, they behave like epidemic threshold models, in the sense that
an infinitesimal initial seed can spread.

Our generalized model therefore exhibits behavior that falls into
one of only three universal classes: class I (epidemic threshold), class II
(vanishing critical mass), and class III (critical mass).  As we show
below, more complicated fixed point curves exist (i.e., curves possessing two or more saddle-node
bifurcation points) but nevertheless
belong to one of these three universal classes, since together they include all
possible behaviors of the fixed point curves near $p=p_c$.

In addition to identifying three universal classes of behavior, it is
also possible to specify the conditions that govern into which class
any particular choice of model parameters will fall. For $r=1$, we can
calculate when the transitions between universal classes occur in
terms of the parameters of interest; that is, the $\{P_k\}$ (viz $f$ and $g$) and
$T$. This exercise amounts to locating the transcritical bifurcation
and determining when it collides with the saddle-node bifurcation.
To locate the transcritical bifurcation, we examine the fixed point equation as $\phifix
\rightarrow 0$.  Since, from \Req{gcontlong.eq:phifixgendosedstar},
\begin{equation}
  \label{gcontlong.eq:phifix->0}
  \phifix = 
  T p \phifix P_1
  + O(\phifix^2),
\end{equation}
we have
\begin{equation}
  \label{gcontlong.eq:transc}
  p_c = \frac{1}{T P_1}.
\end{equation}
The position of the transcritical bifurcation is therefore determined
by the memory length $T$ and the fraction of individuals who are
typically infected by one exposure, $P_1$.  We see immediately that
the transition between class II and class III contagion models occurs
when $p_c=1$; that is, when
\begin{equation}
  \label{gcontlong.eq:transII-III}
  P_1 = 1/T.
\end{equation}

Recalling the homogeneous case, we see that when $\bar{\dstar} = d =1$, $P_1=1$,
giving $p_c=1/T$ as before. It necessarily follows that when
$\bar{\dstar}>d=1$, $P_1=0$ and therefore $p_c=\infty$, thus confirming our
earlier finding that the homogeneous model with $\bar{\dstar}>1$ is always
in the pure critical mass class (i.e., the lower, unstable fixed point
curve of the saddle node bifurcation must have $\phifix>0$ for all
$p_b \le p \le 1$ since it only reaches $\phifix=0$ in the
limit $p \rightarrow \infty$).  We also see that technically all models possess
a transcritical bifurcation somewhere along the $p$-axis, even though
it may be located beyond $p_c>1$ (when $P_1=0$, it lies at $p_c=\infty$).

In order to locate the transition between classes I and II, we determine
when the transcritical and saddle-node bifurcations are coincident,
i.e., when $\dee{\phifix}/\dee{p}=\infty$ at $(p,\phifix)=(p_c,0)$.  
In other words, we calculate
when the fixed point curve emanating from $p=p_c$ is at the transition
between having a large positive slope (class I) and a large negative
slope (class II). We find the condition for this first transition is
\begin{equation}
  \label{gcontlong.eq:transIcond}
  P_1 = P_2/2,
\end{equation}
where details of this calculation are provided in Appendix~\ref{gcontlong.app:transI-IIsec}.

The condition of \Req{gcontlong.eq:transIcond} is a
statement of linearity in the $\{P_k\}$, though importantly only for $k=1$ and $k=2$.
Providing $p_c<1$, if $P_1 > P_2/2$ (i.e., sublinearity holds) a
contagion model is class I whereas if $P_1 > P_2/2$ (i.e.,
superlinearity holds) it is class II.  This condition means that class II
contagion models arise when, on average, two doses are more than twice
as likely as one dose to cause infection.

When this condition is satisfied exactly,
the system's phase transition is a
continuous one as per those of class I, but of a different
universality class: class I systems exhibit a linear scaling near the
critical point whereas the scaling when $P_1=P_2/2$ is $\phifix
\propto (p-p_c)^{1/2}$ (in Appendix~\ref{gcontlong.app:transI-IIsec},
we show that a sequence of increasingly specific
exceptions to this scaling exist depending on the extent of linearity present in the $\{P_k\}$).  

Thus, for $r=1$, the condition that determines whether a given
system is described by an epidemic threshold model, a critical mass model, or a member of 
the intermediate class of vanishing critical mass models, 
depends only on $T$, $P_1$, and $P_2$---a surprising result
given that \Req{gcontlong.eq:phifixgendosedstar} clearly
depends on all the $\{P_k\}$.
A summary of the three basic system types, the transitions
between them, and the accompanying conditions is given in
Table~\ref{gcontlong.tab:systemsummary}.

\begin{table}[tbp]
  \centering
  \begin{tabular}{l|l}
    \hline
    Model class: & Conditions: \\
    \hline
    I: Epidemic threshold &
    $P_1 > P_2/2$ and $P_1 > 1/T$ \\
    I-II transition & 
    $P_1 = P_2/2$ and $P_1 > 1/T$ \\
    II: Partial critical mass & 
    $P_1 < P_2/2$ and $P_1 > 1/T$ \\
    II-III transition & 
    $P_1 < P_2/2$ and $P_1 = 1/T$ \\
    Pure critical mass &
    $P_1 < P_2/2$ and $P_1 < 1/T$ \\
    \hline
  \end{tabular}
  \caption{
    Summary of basic states of the $r=1$, heterogeneous version of model along
    with the corresponding parameter ranges.
    The $P_k$ quantities depend on the distributions of dose size
    and individual thresholds, and are given in~\Req{gcontlong.eq:phifixgendosedstar0}.
  }
  \label{gcontlong.tab:systemsummary}
\end{table}

\begin{figure}[tbp!]
  \centering
  \includegraphics[width=0.49\textwidth]{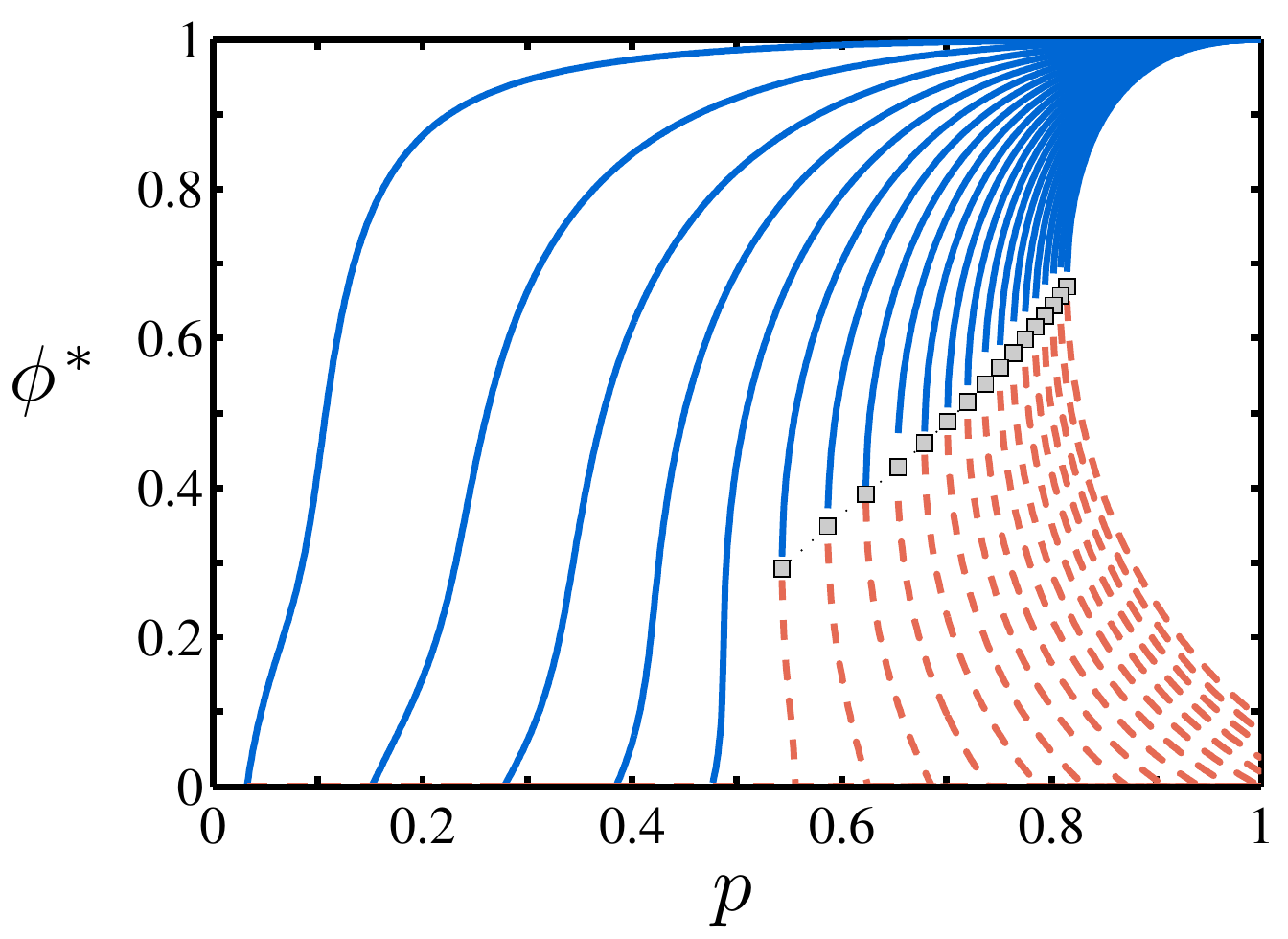} 
  \caption{
    Example of transitions between all three classes of
    contagion models occurring as the probability of recovery $r$ changes.
    Shown are theoretically derived fixed point curves for a range of $r$
    for $T=3$ and a heterogeneous population with threshold distribution 
    $g(\dstar) = 0.3 \delta(\dstar-1) + 0.7 \delta(\dstar-2)$.
    From left to right, the curves correspond to $r=0.01, 0.05, 0.10, 0.15, \ldots, 1.00$.
    Solid lines indicate stable fixed points,
    dashed lines unstable ones, and the squares mark saddle-node bifurcation points.
    The $p$-axis is a line of fixed points,
    and for all fixed point curves, dashed lines indicate unstable fixed points
    and solid lines stable ones. 
    For $r>0.75$, the system belongs to class III (see \Req{eq:gcontlong.eq:transII-IIIr}).
    Fixed points lying on the $p$-axis are not indicated in the plot.
    For each instance of the model, points on the $p$-axis
    to the left of $p_c$ (the intercept of the non-zero fixed point curve with
    the $p$-axis) are stable and points to the right are unstable.
    The expression used here to determine the fixed points is formed by
    appropriately weighting a combination of
    Eqs.~\req{gcontlong.eq:r<1k1T>1}, \req{gcontlong.eq:Gamma2}, and \req{gcontlong.eq:Gammafull}.
  }
  \label{gcontlong.fig:gc_T3_k2k1_bif_theorycomp}
\end{figure}

For $r<1$, the position of the transcritical bifurcation,
\Req{gcontlong.eq:transcr}, generalizes in the same manner as
\Req{gcontlong.eq:r<1k1T>1pc2}. We find
\begin{equation}
  \label{gcontlong.eq:transcr}
  p_c = \frac{1}{P_1(T + \tau)},
\end{equation}
where we recall that $\tau = 1/r - 1$.  As we will see 
in Section~\ref{gcontlong.sec:SIR}, the above statement is
also true for $\RtoSprob<1$ and thus stands as a completely general
result for the model.  We therefore have a condition for the
transition between class II and class III contagion models for a given
$T$ and $r$, analogous to that for the $\rho=1$ case
in~\Req{gcontlong.eq:transII-III}: $P_1 = 1/(T+\tau)$.  For $r<1$, the
condition for the transition between class I and class II contagion
models is more complicated both in derivation and form.  We observe
that as $r$ is decreased, class III models must at some point
transition to class II models and class II models
will eventually become class I models, where we can 
determined the former tranisition by setting
$p_c=1$ in \Req{gcontlong.eq:transcr} and solving for $r$:
\begin{equation}
  \label{eq:gcontlong.eq:transII-IIIr}
  r = \frac{P_1}{1+P_1 - P_1T}.
\end{equation}
In \Req{eq:gcontlong.eq:transII-IIIr} we have assumed $P_1>0$ and $P_1T<1$, since the $r=1$ limit is, by assumption, 
a class III contagion model.
Fig.~\ref{gcontlong.fig:gc_T3_k2k1_bif_theorycomp} presents an
example of a model transitioning from Class III through Class II to Classs I.

% Composite fixed point diagrams

\subsection{Composite classes of dynamics}
\label{gcontlong.subsec:composite}

\begin{figure}[tbp!]
  \centering
  \includegraphics[width=0.49\textwidth]{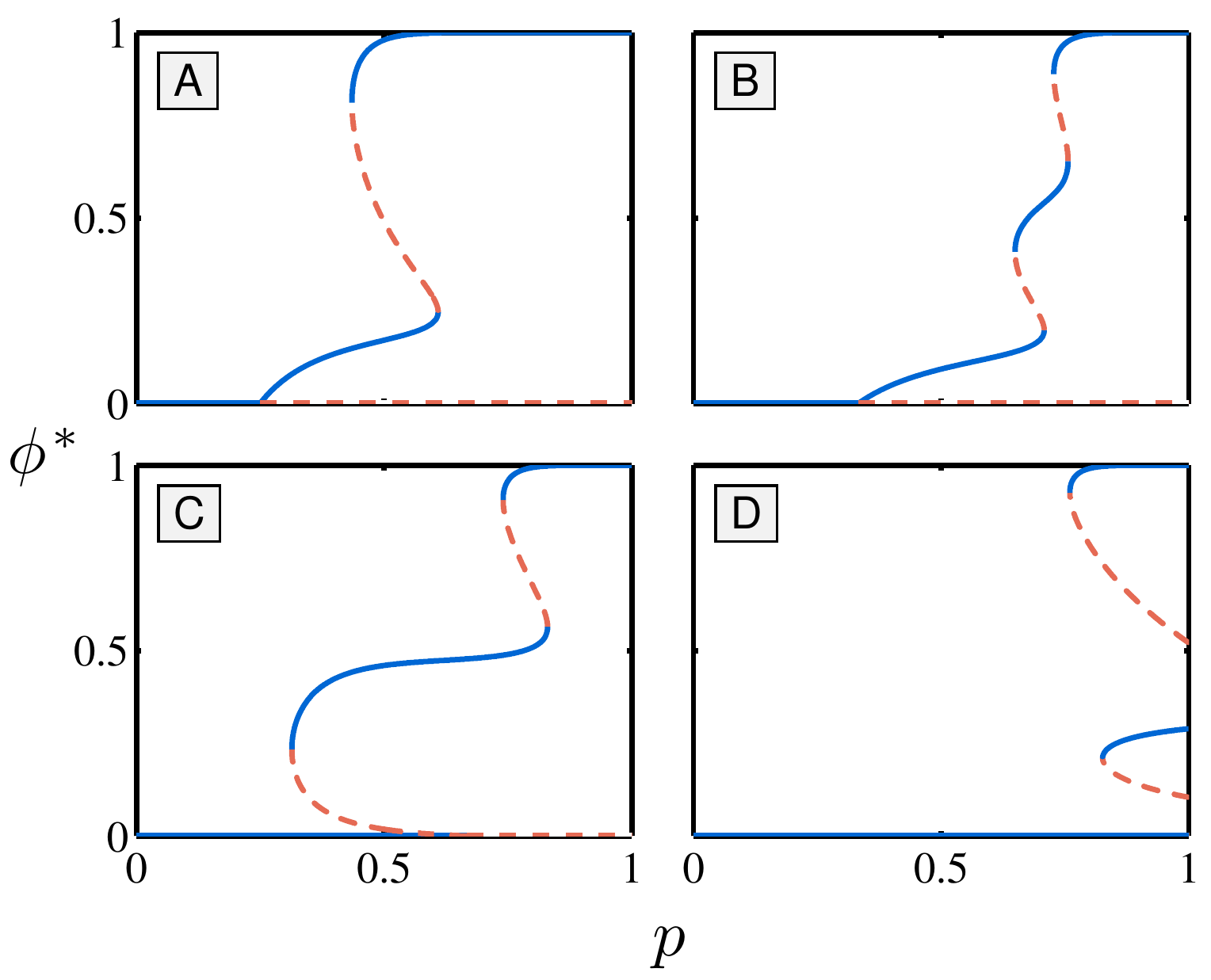} 
  \caption{
    Fixed point curves involving multiple bifurcations.  In all four examples,
    $T=20$ , $r=1$, and dose size is uniformly held at unity.
    The different curves are obtained by adjusting the threshold distribution
    $g$ which in turn leads to changes in the $\{P_k\}$ [see~\Req{gcontlong.eq:phifixgendosedstar0}].
    The curves correspond to 
    (A) $g(d) = 0.2\delta(d-1) + 0.8\delta(d-6)$;
    (B) $g(d) = 0.15\delta(d-1) + 0.4\delta(d-5) + 0.45\delta(d-12)$;
    (C) $g(d) = 0.075\delta(d-1) + 0.4\delta(d-2) + 0.525\delta(d-12)$;
    and
    (D) $g(d) = 0.3\delta(d-3) + 0.7\delta(d-12)$.
    Note that example (D) consists of two separate fixed point curves.
    The curves were found numerically by solving for $p(\phifix)$ 
    using~\Req{gcontlong.eq:phifixgendosedstar}.  
    Solid and dashed lines indicate stable and unstable fixed points respectively.
  }
  \label{gcontlong.fig:manybif}
\end{figure}

Although our main results (three universal classes of behavior and
the conditions that govern the transitions between the three) involve
examples of contagion models with at most two bifurcations, other more
complicated kinds of equilibrium behavior are possible.
Fig.~\ref{gcontlong.fig:manybif} shows four examples of what can
happen for particular distributions of $\dstar$ across the
population. In each example, $T=20$ and $r=1$, all doses are of unit
size, and the population is divided into either two or three
subpopulations with distinct values of $\dstar$. The main features of
each system are captured by the number and locations of the
saddle-node bifurcations. As was the case for the homogeneous model [see
\Req{gcontlong.eq:r=1T>1k>1diff2}], we are able to find an expression
for $z=p_b\phifixb$:
\begin{equation}
  \label{gcontlong.eq:gdiff_eq}
  0 
   =  \sum_{k=1}^{T}
  \binom{T}{k}
  P_k
  z^{k-2}
  (1-z)^{T-k-1}
  [k-1 -z(T-1)],
\end{equation}
where the details of this calculation are provided in
Appendix~\ref{gcontlong.app:saddlenodeextra}. In principle,
\Req{gcontlong.eq:gdiff_eq} could be analyzed to deduce which
$\{P_k\}$ (and hence which $f$ and $g$) lead to
what combination of bifurcations.  
While substantially more complicated, and thus beyond the
scope of this paper, such a classification scheme would be
a natural extension of our
present delineation of the model into three universal classes of
contagion models, based on the behavior near $p=p_c$.  One simple
observation, however, is that \Req{gcontlong.eq:gdiff_eq} is a polynomial of
order $T-2$ and hence a maximum of $T-2$ saddle-node bifurcations may
exist.  From our investigations this outcome seems unlikely as nearby
bifurcations tend to combine with or overwhelm one another;
in other words, subpopulations with sufficiently distinct $\dstar$ are required to
produce systems with multiple saddle-node bifurcations.  Furthermore,
the distribution of dose sizes $f$ seems unlikely to be
multimodal for real contagious influences or entities,
and the way it enters into the calculation of the $\{P_k\}$
[\Req{gcontlong.eq:phifixgendosedstar0}] reduces its effect in
producing complicated systems.  Thus, the number of distinct bifurcations
is limited and strong multimodality in the threshold distribution $g$
appears to be the main mechanism for producing systems with more
than one saddle node bifurcation.
As a first step in this extended analysis of the model, we derive 
in Appendix~\ref{gcontlong.app:saddlenodeextra} the condition
for the appearance of two saddle-node bifurcations (i.e., one forward
and one backward).

\section{SIRS and SIR Contagion Models}
\label{gcontlong.sec:SIR}

As mentioned in section~\ref{gcontlong.sec:model}, the SIS
class of behavior that we have analyzed exclusively up to now is a
somewhat special case of the general contagion process as it assumes
that recovered individuals instantly become re-susceptible.  This
assumption renders the SIS case particularly tractable, and we have
taken advantage of this fact in the preceding sections to make some
headway in understanding the full range of equilibrium behavior of the
model.  However, it remains the case that very few, if any, infectious
diseases could be considered to obey true SIS-type dynamics, as almost all
recovery from infection tends to be associated with some finite period
of immunity.  Any purportedly ``general'' model of contagion ought
therefore to be analyzed in a wider domain of the associated removal
period, and any corresponding classes of behavior labeled ``universal''
ought to withstand the introduction of at least some period of
immunity to re-infection. Thus motivated, we now extend our previous
analysis to systems where individuals experience temporary (SIRS, $0 <
\RtoSprob < 1$) or permanent removal (SIR, $\RtoSprob=0$). We present
some preliminary results for each of these cases in turn, relying now exclusively on 
numerical simulation.

\begin{figure}[tbp]
  \centering
  \includegraphics[width=0.49\textwidth]{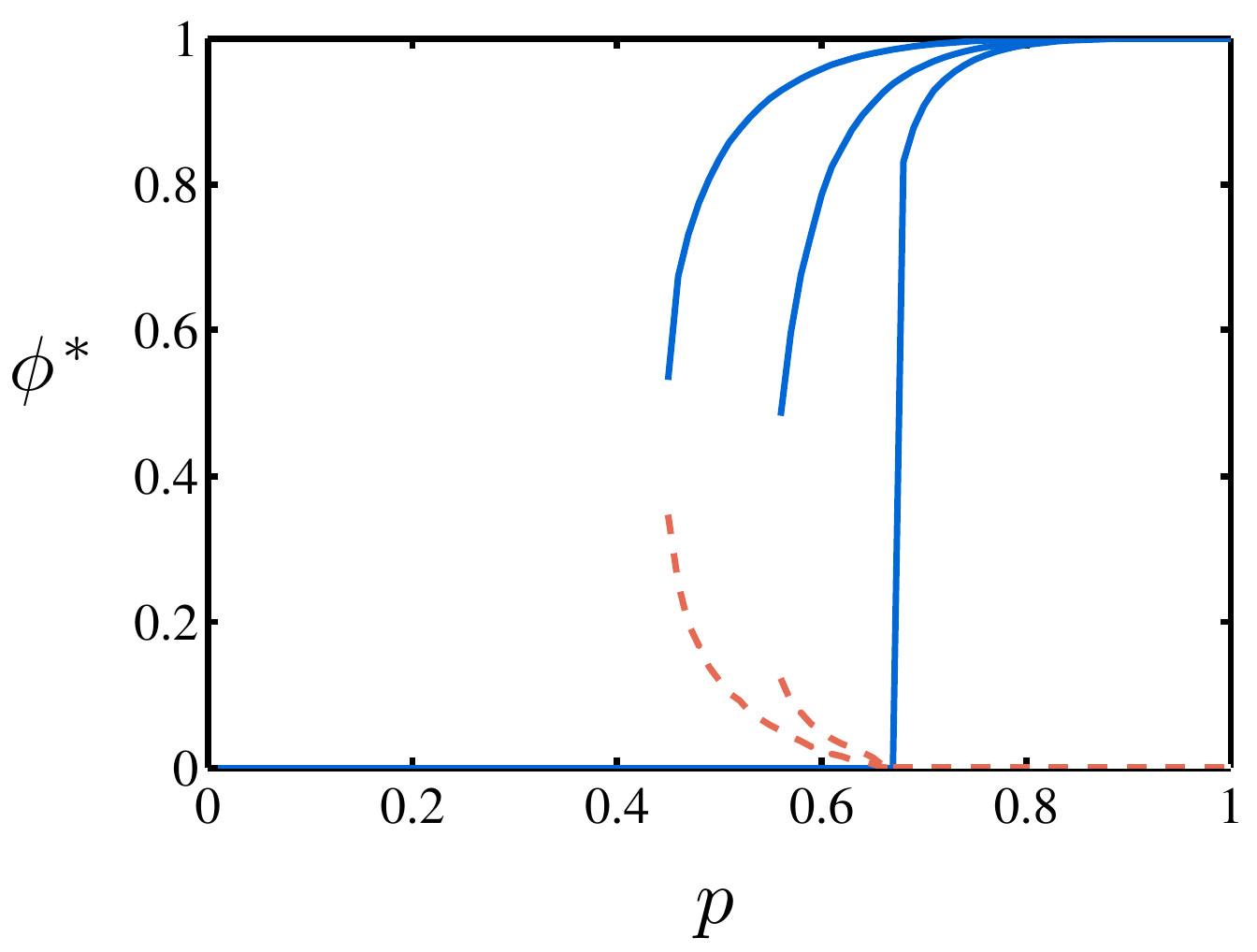}
  \caption{
    The effect of reducing $\RtoSprob$ (the probability of an immune individual becoming
    susceptible) for a contagion model that is class II when $\RtoSprob=1$.
    Here, $N=10^5$, $T=6$, $r=1$, dose sizes are fixed at unity,
    and $g(\dstar) = 0.25 \delta(\dstar-1) + 0.75 \delta(\dstar-2)$.
    From left to right, the plots correspond to 
    $\RtoSprob=1$, $0.15$, and $0.09$.
    In all cases, $p_c$, the position of the $p$-axis intercept, is independent of $\RtoSprob$.
    Apart from the intercept at $p=p_c$ and the point $(p,\phifix)=(1,1)$, all other points of the curve move
    to the right as $\RtoSprob$ decreases.
    Consequently, as $\RtoSprob$ is reduced,
    all class II contagion models will at some point become members of class I.
    Contagion models in class I and class III remain unchanged in their nature.
    In finding the upper stable branch of $\phifix(p)$, all individuals are initially infected.
    A binary search is then used to detect the position of the lower unstable branch.
  }
  \label{gcontlong.fig:rhoclasstrans}
\end{figure}

For SIRS contagion, we observe that the position of the transcritical
bifurcation $(p_c,0)$ does not change as $\RtoSprob$ is reduced from 1;
however, all non-zero fixed points move in the positive $p$ direction.  
Because they remain in the removed state for a longer time, individuals in systems with lower $\RtoSprob$
spend relatively less time infected than those in systems with higher $\RtoSprob$
(this is in contrast to the effect of reducing $r$, 
which prolongs the time individuals are infected,
thereby causing fixed points
to move in the negative $p$ direction).
Thus, contagion models
belonging to class I and class III in the $\RtoSprob=1$ special case
remain in their respective  classes as $\RtoSprob$ decreases.  
However, as shown in Fig.~\ref{gcontlong.fig:rhoclasstrans},
class II models will transition to class I for some $\RtoSprob < 1$.

% 3. rho=0 case: describe persistence

\begin{figure}[tbp]
  \centering
  \includegraphics[width=0.49\textwidth]{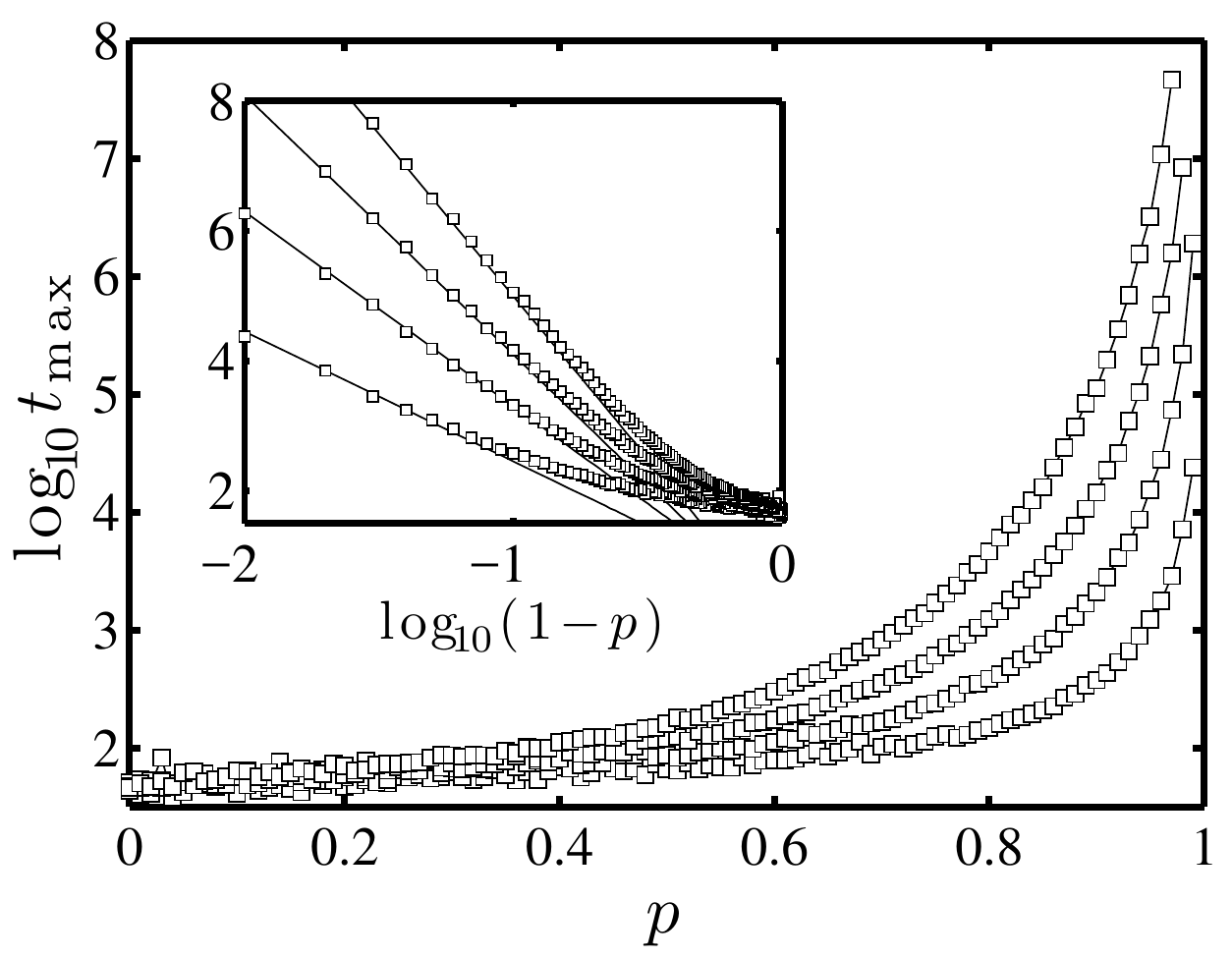}
  \caption{
    Time taken for an initially universal infection to die out in the SIR case ($\RtoSprob=0$).
    Moving from bottom to top in both plots, $T=3$, 4, 5, and 6.
    The model parameters used here are $\kstar=1$, unit dose size, $r=0.2$,
    and population size $N=10^4$.   For all systems, $p=\phi=1$ is a
    fixed point and hence $t_{\rm max}=\infty$ at $p=1$.
    The main plot shows the rapid increase in $t_{\rm max}$ for $p \rightarrow 1$.
    The inset shows the behavior of $t_{\rm max}$ as a function of  $(1-p)$
    as plotted on a double logarithmic scale.
    The lines have slopes of $-(T-1)$ indicating that infections are
    strongly persistent with $\tmax \propto (1-p)^{-(T-1)}$.
  }
  \label{gcontlong.fig:SIR_logn_050}
\end{figure}

For the SIR version of the model, recovered individuals cannot return
to the susceptible state, and no fraction of the population remains
infected in the infinite time limit; hence we can no longer speak of
non-zero fixed point curves, and the fraction of individuals infected and
recovered, $\phi(t)$ and $R(t)$, become
the relevant objects of study.  Some defining quantities are then
the maximum fraction of the population infected at any one time, $\max_t \phi(t)$, the fraction
eventually infected, $1-\phi(\infty)-R(\infty)$, and the \textit{relaxation time}
required for the epidemic
to die out.  We focus on the latter here which we denote by $\tmax$.

We observe that when individuals possess a memory of doses (i.e., $T>1$), $\tmax$
diverges as $p \rightarrow 1$.  
Figure~\ref{gcontlong.fig:SIR_logn_050} shows $\tmax(p)$
for four sample systems with $T=3$, 4, 5, and 6.  We find the
divergence of $\tmax$ near $p=1$ to be well approximated by
\begin{equation}
  \label{gcontlong.eq:SIR_tmax}
  \tmax \propto (1-p)^{-(T-1)},
\end{equation}
where fits are shown in the inset of Fig.~\ref{gcontlong.fig:SIR_logn_050}.
Equation \req{gcontlong.eq:SIR_tmax} also 
shows that for a fixed $p$, the relaxation time increases exponentially with 
length of memory $T$ for all $p$, i.e., 
\begin{equation}
  \label{gcontlong.eq:SIR_tmax2}
  \tmax \propto e^{\alpha T},
\end{equation}
where $\alpha=-\ln{(1-p)}>0$.

Thus, when $\RtoSprob=0$, epidemics,
while not ever achieving a non-zero steady state as in the $\RtoSprob>0$ case,
can persist for (arbitrarily) long periods of time.
The introduction of memory, which allows
infected individuals to maintain their dose count 
above their threshold by repeatedly infecting 
each other, creates an SIR model with strikingly different
behavior to the standard memoryless SIR model.

\section{Concluding remarks}
\label{gcontlong.sec:concl}

Our aim here has been to develop and analyze in detail a model of
contagion that incorporates and generalizes elements of contagion models from
both the social sciences and epidemiology.
A key feature of the current model is
that interdependencies between successive exposures are introduced in
a natural way by varying the length of memory that an individual
maintains of past exposures.  Contagion models incorporating memory correspond to
standard notions of contagion in the social sciences (although these
models rarely discuss the role of memory explicitly), while
memory-less contagion models correspond to traditional models of
disease spreading.  Our model suggests, however, that in reality these
two kinds of contagion models may not be entirely distinct.  We
venture the possibility that some infectious diseases may spread in a
fashion similar to social contagion processes.  For example,
two exposures to an agent sufficiently close in time may infect an
individual with higher probability than would be 
expected if the exposures acted as independent events.
The response of some individuals to
allergens where separate doses accumulate in the body may be an
example of a disease with memory.  Although allergies are not
contagious, they demonstrate that such a phenomenon is biologically
plausible---a possibility that has largely remained unexamined in the
microorganismal dose-response literature~\citep{haas2002}.

The main result of our analysis is the identification of three
universal classes of contagion dynamics, along with precise conditions
for the transitions between these classes.  Given the
complexity of the model, these conditions are surprisingly simple---at
least in the SIS case ($r=\RtoSprob=1$, see
Table~\ref{gcontlong.tab:systemsummary})---and present us with
quantities such as $T$, $P_1$, and $P_2$ that may in principle be
measurable for real epidemics.  Furthermore, the dependence of the
transition conditions only on $P_1$, $P_2$, and $T$, rather than on the full
details of the underlying distributions of thresholds $(g(\dstar))$
and doses $(f(d))$, suggests a new and possibly useful level of
abstraction for thinking about contagious processes; that is,
measuring individuals in terms of their dose-response and
characterizing a population in terms of its $\{P_k\}$.

For the more complicated and general cases of $r<1$ (finite recovery
period) and $\RtoSprob<1$ (finite immunity period), we have confirmed that
the same basic three-class structure persists, and determined the
position of the transcritical bifurcation $p_c$,
\Req{gcontlong.eq:transcr}, that is one of the two quantities needed
to specify the conditions for transitioning between classes. The other
condition, derived from calculating the slope of the fixed point curve
as it passes through the transcritical bifurcation, merits further
attention.  All of these conditions, however, are ultimately dependent
on the distributions $f$ and $g$.  Our analysis of the model suggests,
for example, that composite fixed point diagrams involving more than
one saddle point node can only result from multimodality in $g$, the
distribution of thresholds. Exactly how the details of these two
distributions affect $P_1$ and $P_2$ would be worth further
investigation.

Our model suggests that some epidemics may be prevented or enabled
with slight changes in system parameters (if feasible). For example,
knowing that a potentially contagious influence belongs to class II and that $p$
is just below $p_c$ would indicate that by increasing inherent
infectiousness (i.e., $p$) or by creating a sufficiently large enough
base of infected individuals, the contagion could be kicked off with
potentially dramatic results. Alternatively, by increasing $r$ or
reducing $T$ or $\RtoSprob$, the possibility of undesirable epidemics may be reduced,
as for all these adjustments fixed point curves are generally moved in the
direction of higher values of $p$.

Finally, we have focused exclusively in this paper on a
 mean field analysis of the model, which
is to say, we have made the standard
assumption that individuals in a population mix uniformly and at
random.  A natural generalization would
be to consider the model's behavior for a networked population of
individuals.  Other simulation possibilities would be to consider
distributions of $T$, $p$, $r$, and $\RtoSprob$. Finally, additional technical
investigation of this model would also include analysis of the 
closed-form expression for saddle-node bifurcations given in
\Req{gcontlong.eq:gdiff_eq}, and a derivation of analytic expressions
for the model when $\RtoSprob<1$ for small $T$ and $\dstar$.  We hope that
our preliminary investigations into this interesting and reasonably
general class of contagion models will stimulate other researchers to
pursue some of these extensions.

\acknowledgments

We are grateful for discussions with Duncan Callaway, Daryl Daley, Charles Haas,
and Matthew Salganik.
This research was supported in part by the National
Science Foundation (SES 0094162), the Office of Naval Research, Legg
Mason Funds, and the James S.\ McDonnell Foundation.

\appendix

\section{Conditions for existence of saddle-node bifurcation}
\label{gcontlong.app:saddlenodeextra}

In \Req{gcontlong.eq:gdiff_eq}, we provided a
closed form expression for $z=p_b\phifixb$ where
$(p_b,\phifixb)$ is the location of a saddle-node bifurcation.  
This expression pertains to the heterogeneous
version of the model for $r=\RtoSprob=1$ [the homogeneous version 
is given in \Req{gcontlong.eq:r=1T>1k>1diff2}].
As noted in the main text, this equation has up to $T-2$ solutions,
depending on the form of the $\{P_k\}$.
In this Appendix, we derive \Req{gcontlong.eq:gdiff_eq}
and also find a criterion for the appearance of two
saddle-node bifurcations.  All calculations revolve
around determining when the slope of the fixed point
curve $\phifix(p)$ becomes infinite, or, equivalently,
finding when $\tdiff{p}{\phifix}=0$. 

Our starting point is \Req{gcontlong.eq:phifixgendosedstar},
the general closed form expression for $\phifix$ as a function of $p$,
from which we can calculate $\tdiff{p}{\phifix}$.
We rewrite \Req{gcontlong.eq:phifixgendosedstar} as
\begin{eqnarray}
  \label{gcontlong.eq:g_phifixgendosedstar}
  \hfun(p,\phifix)
  & = &
  -\phifix 
  + 
  \sum_{k=1}^{T}
  \binom{T}{k}
  P_k
  (p\phifix)^{k}
  (1-p\phifix)^{T-k}, 
  \nonumber \\
  & = &
  \phifix \left[
    -1 
    + 
  p \sum_{k=1}^{T}
  \binom{T}{k}
  P_k
  (p\phifix)^{k-1}
  (1-p\phifix)^{T-k} \right], 
  \nonumber \\
  & = &
  \phifix \, \Hfun(p, \phifix),
\end{eqnarray}
with the requirement $\hfun(p,\phifix)=0$.

We show that we can use the function
$\Hfun$ instead of $\hfun$ to find $\tdiff{p}{\phifix}$
when $\phifix \ne 0$.
Differentiating $\hfun(p,\phifix) = \phi \Hfun(p, \phifix) = 0$ with
respect to $\phifix$, we have
\begin{equation}
  \label{gcontlong.eq:hphiHdiff}
  \pdiff{\hfun}{\phifix}
  + \diff{p}{\phifix}
  \pdiff{h}{p}
  = 
  \Hfun
  +
  \phifix \pdiff{\Hfun}{\phifix}
  + 
  \phifix 
  \diff{p}{\phifix}
  \pdiff{\Hfun}{p} = 0.
\end{equation}
Again, since $\hfun(p,\phifix)=\phifix \Hfun(p,\phifix)=0$, we have $\Hfun(p,\phifix)=0$ when
$\phifix \ne 0$.   When we also
require $\tdiff{p}{\phifix}=0$ (i.e., the chief condition for a 
saddle-node bifurcation point),
\Req{gcontlong.eq:hphiHdiff} reduces to
\begin{equation}
  \label{gcontlong.eq:hphiHdiff2}
  \pdiff{\Hfun}{\phifix}
  = 0,
\end{equation}
and so we may find solutions of $\tpdiff{\Hfun}{\phifix}=0$
instead of $\tpdiff{\hfun}{\phifix}=0$.
The benefit of making this observation is that we find
$\tpdiff{\Hfun}{\phifix}$ can be expressed in terms of a single variable ($z = p_b\phifixb$),
allowing for simpler analytical and numerical examination (recall that $(p_b,\phifixb)$ 
denotes the position of a saddle-node bifurcation).
Returning to the definition of $\Hfun$ given in \Req{gcontlong.eq:g_phifixgendosedstar},
we find
\begin{eqnarray}
  \label{gcontlong.eq:Hdiffp}
  \lefteqn{\pdiff{\Hfun}{\phifix} = 
  \sum_{k=1}^T \binom{T}{k} P_k p^k (k-1) (\phifix)^{k-2} (1-p\phifix)^{T-k}} \nonumber \\
  & & + \sum_{k=1}^T \binom{T}{k} P_k p^k       (\phifix)^{k-1} (T-k) (-p) (1-p\phifix)^{T-k-1} \nonumber \\
  & & + \diff{p}{\phifix} \left( \ldots \right). 
\end{eqnarray}
Since we require $\tdiff{p}{\phifix}=0$, the above simplifies to
\begin{eqnarray}
  \label{gcontlong.eq:Hdiffp2}
  \lefteqn{\pdiff{\Hfun}{\phifix} = \sum_{k=1}^T \binom{T}{k} P_k p^k (\phifix)^{k-2} (1 -p\phifix)^{T-k-1} }
  \nonumber \\
  &  & \times \left[ (k-1)(1-p\phifix) - p\phifix(T-k) \right]. 
\end{eqnarray}
Setting $\tpdiff{\Hfun}{\phifix}=0$ and removing a factor of $p^2$,
we find the positions of all saddle-node bifurcation points satisfy 
\begin{equation}
  \label{gcontlong.eq:gdiff_eq_app}
  0 =  \sum_{k=1}^{T}
  \binom{T}{k}
  P_k
  z^{k-2}
  (1-z)^{T-k-1}
  [k-1 -z(T-1)],
\end{equation}
where $z = p_b\phifixb$.
Upon solving \Req{gcontlong.eq:gdiff_eq_app} for $z$
(where for all nontrivial solutions, we require $0 < z < 1$),
\Req{gcontlong.eq:phifixgendosedstar}
can then be used to find $\phifixb$ (since it expresses
$\phifixb$ as a function of $z$) and hence $p_b$.

In order to determine whether a bifurcation
is forward or backward facing
(by forward facing, we mean branches emanate
from the bifurcation point in the direction of the positive $p$-axis),
we need to compute $\tdiffsq{p}{\phifix}$,
and examine its sign.
(When $\tdiffsq{p}{\phifix}=0$, 
two saddle-node bifurcations points are coincident,
one forward and one backward facing.)
If the $\{P_k\}$ are parametrized in some fashion
(i.e., $f$ and/or $g$ are parametrized),
then we can determine the relevant parameter 
values at which pairs of bifurcations appear.
We compute an expression for $\tdiffsq{p}{\phifix}$ as follows.

Differentiating~\Req{gcontlong.eq:hphiHdiff} 
with respect to $\phifix$, we have
\begin{equation}
  \label{gcontlong.eq:hphiHdiffd2}
  \diffsq{\hfun}{\phifix}
  =
  2\diff{\Hfun}{\phifix}
  +
  \phifix
  \diffsq{\Hfun}{\phifix} = 0.
\end{equation}
We already have 
$\tdiff{\Hfun}{\phifix}=\tpdiff{\Hfun}{\phifix} + \tdiff{p}{\phifix} \tpdiff{\Hfun}{p} = 0$,
and so~\Req{gcontlong.eq:hphiHdiffd2} now gives
\begin{equation}
  \label{gcontlong.eq:hphiHdiffd2_2}
  \diffsq{\Hfun}{\phifix} = 0.
\end{equation}
Expanding this, we have
\begin{equation}
  \label{gcontlong.eq:hphiHdiffd2_3}
  0 = \pdiffsq{\Hfun}{\phifix} 
  + \diff{p}{\phifix} 
  \frac{\partial^2 \Hfun}{\partial p \partial \phifix}
  + \diffsq{p}{\phifix} \pdiff{\Hfun}{p}
  + \diff{p}{\phi} \frac{\dee}{\dee p} \pdiff{\Hfun}{p}.
\end{equation}
The second and fourth terms on the right hand
disappear since $\tdiff{p}{\phifix}=0$, leaving
\begin{equation}
  \label{gcontlong.eq:hphiHdiffd2_4}
  0 = \pdiffsq{\Hfun}{\phifix} 
  + \diffsq{p}{\phifix} \pdiff{\Hfun}{p}.
\end{equation}
Upon rearrangement, we have
\begin{equation}
  \label{gcontlong.eq:hphiHdiffd2_5}
  \diffsq{p}{\phifix} = 
  - \frac{\tpdiffsq{\Hfun}{\phifix}}
  {\tpdiff{\Hfun}{p}}.
\end{equation}
We first compute $\tpdiff{\Hfun}{p}$.  With
$\Hfun$ as defined in \Req{gcontlong.eq:g_phifixgendosedstar},
we see that
\begin{eqnarray}
  \label{gcontlong.eq:Hpdiff}
  \lefteqn{\pdiff{\Hfun(p,\phifix)}{p} =
  \pdiff{}{p} \frac{1}{\phifix}
  \sum_{k=1}^{T}
  \binom{T}{k}
  P_k (p\phifix)^k(1-p\phifix)^{T-k},} \nonumber \\
  & = &
  \pdiff{}{p\phifix} 
  \sum_{k=1}^{T}
  \binom{T}{k}
  P_k (p\phifix)^k(1-p\phifix)^{T-k}, \nonumber \\
  & = &
  \pdiff{}{z} 
  \sum_{k=1}^{T}
  \binom{T}{k}
  P_k z^k (1-z)^{T-k}, \nonumber \\
  & = &
  \sum_{k=1}^{T}
  \binom{T}{k}
  P_k z^k (1-z)^{T-k-1}
  [ k(1-z) - (T-k)z], \nonumber \\
  & = &
  \sum_{k=1}^{T}
  \binom{T}{k}
  P_k z^k (1-z)^{T-k-1}
  [k - Tz].
\end{eqnarray}
Using Eqs.~\req{gcontlong.eq:g_phifixgendosedstar}
and~\req{gcontlong.eq:gdiff_eq_app}
(i.e., that we are at a bifurcation point),
the above yields
\begin{eqnarray}
  \label{gcontlong.eq:Hpdiff2}
  \pdiff{\Hfun(p,\phifix)}{p} = \frac{1}{p}.
\end{eqnarray}
Using this result, \Req{gcontlong.eq:hphiHdiffd2_5} becomes
\begin{equation}
  \label{gcontlong.eq:hphiHdiff2_2}
  \diffsq{p}{\phifix} = 
  - p\pdiffsq{\Hfun}{\phifix}.
\end{equation}
Next, we find
\begin{eqnarray}
  \label{gcontlong.eq:Hphidiffsq}
  \lefteqn{\pdiffsq{\Hfun}{\phifix}
    = 
    \sum_{k=1}^{T}
    \binom{T}{k}
    P_k z^{k-3} (1-z)^{T-k-2} \times } \\
  & & 
  [(k-1)(k-2) - 2z(k-1)(T-2) + z^2(T-1)(T-2)]. \nonumber
\end{eqnarray}
As stated above, if the $\{P_k\}$ are parametrized
in some way, and we are interested in finding at what
parameter values two bifurcations appear and 
begin to separate, then we need to determine when
$\tpdiff{\Hfun}{\phifix}=0$ [\Req{gcontlong.eq:gdiff_eq_app}]
and when
$\tpdiffsq{\Hfun}{\phifix}=0$ [\Req{gcontlong.eq:Hphidiffsq}].

If however the $\{P_k\}$ are fixed and we want to find all bifurcation points
along with whether they are forward or backward bifurcations, then we can 
check the latter by finding the sign of $\tdiffsq{p}{\phifix}=0$ [\Req{gcontlong.eq:hphiHdiff2_2}].
Further analysis of these equations may be possible to find conditions
on $f$ and $g$, and thereby the $\{P_k\}$, that would ensure 
certain types of model behavior.

\section{Exact solution for $r<1$, $\dstar=2$, and $T=2$ and $T=3$}
\label{gcontlong.app:exactdstar=2}

For $r<1$, in general we have individuals whose cumulative dose is below
the threshold $\bar{\dstar}$ but are still infected because they have not
yet recovered.  To find the proportion of individuals in this
category, we must calculate $\Gamma_m$, the fraction of individuals whose
memory count $D$ [number of successfully infecting interactions, \Req{gcontlong.eq:sumT}]
last equaled the threshold $m$ time steps ago and has been
below the threshold since then.  The fraction
of these individuals still infected will be $(1-r)^m$, i.e., those who have
failed to recover at each subsequent time step.  We write
the proportion of infected individuals below the threshold as
\begin{equation}
  \label{gcontlong.eq:Gamma}
  \Gamma(p,\phifix;r,T) = \sum_{m=1}^{\infty} (1-r)^m \Gamma_m(p,\phifix;T).
\end{equation}
Once $\Gamma$ in determined, a closed form expression
for $\phifix(p)$ is obtained by inserting $\Gamma$ into \Req{gcontlong.eq:Gamma2}.
To determine the $\{\Gamma_m\}$, we explicitly construct all allowable
length $m$ sequences of $1$'s and $0$'s such that no subsequence
of length $T$ has $\dstar=2$ or more $1$'s.
The analysis is similar for both the $T=2$ and $T=3$ cases
we consider in this Appendix, and a generalization to all $T$ is possible.
Below, we first show the forms of $\Gamma$ for the two
cases and then provide details of the calculations involved.

For $T=3$, we obtain
\ifthenelse{\boolean{revtexswitch}}{\begin{widetext}}{}
  \begin{eqnarray}
    \label{gcontlong.eq:Gammafull}
    \lefteqn{\Gamma(p,\phifix;r,T) = (p\phifix)^2 (1-p\phifix)^2 \times }  \\
      & &   \biggl(  1 - r + \sum_{m=1}^{\infty} (1-r)^m     
      \left[ 
        \chi^{(3)}_{m-1} + 
        \chi^{(3)}_{m-2} + 
        \right. \biggr. \nonumber \\
 & &      \biggl. \left.
      2p\phifix (1-p\phifix)\chi^{(3)}_{m-3} + 
      p\phifix (1-p\phifix)^2\chi^{(3)}_{m-4}
    \right] \biggr) \nonumber
  \end{eqnarray}
\ifthenelse{\boolean{revtexswitch}}{\end{widetext}}{}
where $\chi^{(T)}_m$ is defined as
\begin{equation}
  \label{gcontlong.eq:chigeneral}
  \chi^{(T)}_m(p,\phifix)
  = 
  \sum_{k=0}^{[m/T]}
  \binom{m-(T-1)k}{k}
    (1-p\phifix)^{m-k}
    (p\phifix)^k.
\end{equation}
Upon inserting $\Gamma$ into \Req{gcontlong.eq:Gamma2},
we have a closed form expression for $\phifix$ involving
$p$ and $r$ as parameters.  This expression can then be solved for numerically
yielding the fixed point curves in Figure~\ref{gcontlong.fig:gc_T3_k2_bif_theorycomp2}.

For the $T=\bar{\dstar}=2$ case, we find
\begin{eqnarray}
  \label{gcontlong.eq:GammaT2k2}
    \lefteqn{\Gamma(p,\phifix;r,T) = }  \\
    & & (p\phifix)^2 (1 - p\phifix)
    \sum_{m=1}^\infty
    (1-r)^m
    \left[
      \chi^{(2)}_{m-1}
      +
      p\phifix
      \chi^{(2)}_{m-2}
    \right]. \nonumber
\end{eqnarray}

We consider the case of $\dstar=2$ and $T=3$ first.
For this specification of the model, there are two ways 
for an individual to transition to 
being below the threshold, i.e., $D<\dstar$.  
An individual must have two positive
signals and one null signal in their memory and then receive a null
signal while losing a positive signal out the back of their
memory window.  The two sequences for which this happens are
\begin{equation}
  \label{gcontlong.eq:T3k2seq1}
  \{d_{n-2},d_{n-1},d_{n},d_{n+1}\} = \{1,1,0,0\}
\end{equation}
and
\begin{equation}
  \label{gcontlong.eq:T3k2seq2}
  \{d_{n-2},d_{n-1},d_{n},d_{n+1}\} = \{1,0,1,0\},
\end{equation}
with the point of transition to being below the threshold
occurring between time steps $n$ and $n+1$.
The two other sequences for which a node will be above
the threshold are $\{1,1,1\}$ and $\{0,1,1\}$ but neither
of these can drop below the threshold of $\dstar=2$ in the next time step.
Both the sequences of Eqs.~\req{gcontlong.eq:T3k2seq1}
and~\req{gcontlong.eq:T3k2seq2} occur with probability
$(1-p\phifix)^2 (p\phifix)^2$.  
When $m>1$, $d_{n+2}$ may be either $0$ or $1$ for the first sequence,
but for the second $d_{n+2}=0$ or otherwise the threshold will be reached again:
\begin{equation}
  \label{gcontlong.eq:T3k2seq2b}
  \{d_{n-2},d_{n-1},d_{n},d_{n+1},d_{n+2}\} = \{1,0,1,0,0\},
\end{equation}

Given these two possible starting points,
we now calculate the number of paths for which
$D_{t+j}$ remains below $\dstar=2$ for $j=1,\ldots,m$.
The structure of an acceptable sequence must be such that
whenever a 1 appears, it is followed by at least
two 0's (otherwise, $\dstar$ will be exceeded).  
We can see therefore that every allowable
sequence is constituted by only two distinct subsequences:
$a=\{0\}$ and $b=\{1,0,0\}$.  

Our problem becomes
one of counting how many ways there are to arrange a
sequence of $a$'s and $b$'s given an overall sequence
length $m$ and that the length of $a$ is 1 and the
length of $b$ is 3.  

If we fix the number of subsequences
of $a$ and $b$ at $N_a$ and $N_b$ then we must have
$m = N_a \cdot 1 + N_b \cdot 3$.  Varying $N_b$ from
$0$ to $[m/3]$ (the square brackets indicate the integer
part is taken), we have $N_a$ correspondingly varying
from $m$ down to $m-3[m/3]$ in steps of $3$.

Next, we observe that the number of ways of arranging $N_a + N_b$
subsequences $a$ and $b$ is 
\begin{equation}
  \label{gcontlong.eq:NaNb}
  \binom{N_a + N_b}{N_a} = \binom{N_a + N_b}{N_b} =  \frac{(N_a+N_b)!}{N_a! N_b!}.
\end{equation}
To see this, consider a sequence of slots labeled $1$ through $N_a+N_b$.
In ordering the $N_a$ $a$'s and $N_b$ $b$'s, we are asking
how many ways there are to choose $N_a$ slots for the $a$'s (or
equivalently $N_b$ slots for the $b$'s).  We are interested in the labels of the slots
but not the order that we select them so we obtain the binomial coefficient
of~\Req{gcontlong.eq:NaNb}.

Allowing $N_b$ and $N_a$ to vary while holding $m = N_a + 3 N_b$ fixed,
we find the total number of allowable sequences to be
\begin{equation}
  \label{gcontlong.eq:chi1}
  \sum_{N_b=0}^{[m/3]}
  \binom{N_b+N_a}{N_b}
  = 
  \sum_{k=0}^{[m/3]}
  \binom{m-2k}{k},
\end{equation}
where we have replaced $N_b$ with $k$ and $N_a$ with  $3N_b-m= 3k-m$.
Noting that the probability of $a$ is $(1-p\phifix)$ and $b$ is $p\phifix(1-p\phifix)^2$,
the total probability $\chi^{(3)}_m(p,\phifix)$ of all allowable sequences of length $m$ for $T=3$ follows
from~\Req{gcontlong.eq:chi1}:
\begin{equation}
  \label{gcontlong.eq:chi2}
  \chi^{(3)}_m(p,\phifix)
%  & = & 
%  \sum_{k=0}^{[m/3]}
%  \binom{m-2k}{k}
%  \left[
%    (1-p\phifix)^2 p\phifix
%  \right]^k
%  (1-p\phifix)^{m-3k} \nonumber \\
  = 
  \sum_{k=0}^{[m/3]}
  \binom{m-2k}{k}
    (1-p\phifix)^{m-k}
    (p\phifix)^k.
\end{equation}
For general $T$, $\chi^{(T)}_m$ is defined
by equation \Req{gcontlong.eq:chigeneral}.

We must also address some 
complications at the start and end of allowable sequences.  
At the end of a sequence, we have the issue
of 1's being unable to appear because our component subsequences
are $a=\{0\}$ and $b=\{1,0,0\}$.  Any sequence ending in two or more $0$'s will
be accounted for already but two endings we need to
include are $\{d_{n+m}\} = \{1\}$ and $\{d_{n+m-1}, d_{n+m}\} = \{1,0\}$.
We do this for each of the two starting sequences and we therefore
have six possible constructions for allowable sequences of length $m$.
For the starting sequence given in \Req{gcontlong.eq:T3k2seq1}, we have
the following three possibilities
for $\{d_{n-2}, d_{n-1}, d_{n}, d_{n+1}, \ldots, d_{n+m}\}$:
\begin{gather}
  \label{gcontlong.eq:allowableseq1}
  H_1 = \{1, 1, 0, 0, H_{m-1}^{a,b}\}, \\
  \label{gcontlong.eq:allowableseq2}
  H_2 = \{1, 1, 0, 0, H_{m-2}^{a,b}, 1\}, \\
  \mbox{and} \nonumber \\
  \label{gcontlong.eq:allowableseq3}
  H_3 = \{1, 1, 0, 0, H_{m-3}^{a,b}, 1, 0\},
\end{gather}
where $H_{m}^{a,b}$ is a length $m$ sequence of $a$'s and $b$'s
[which as we have deduced occur with probability $\chi^{(3)}_{m}(p,\phifix)$].
For the starting sequence given in \Req{gcontlong.eq:T3k2seq2b}, we similarly have
\begin{gather}
  \label{gcontlong.eq:allowableseq4}
  H_4 = \{1, 0, 1, 0, 0, H_{m-2}^{a,b}\}, \\
  \label{gcontlong.eq:allowableseq5}
  H_5 = \{1, 0, 1, 0, 0, H_{m-3}^{a,b}, 1\}, \\
  \mbox{and} \nonumber \\
  \label{gcontlong.eq:allowableseq6}
  H_6 = \{1, 0, 1, 0, 0, H_{m-4}^{a,b}, 1, 0\}.
\end{gather}
The probabilities corresponding to sequences~\req{gcontlong.eq:allowableseq1}
through~\req{gcontlong.eq:allowableseq6} are
\begin{gather}
  \label{gcontlong.eq:allowableseqprob1}
  {\rm Pr}(H_1) = (p\phifix)^2 (1-p\phifix)^2 \chi^{(3)}_{m-1}(p,\phifix), \\
  \label{gcontlong.eq:allowableseqprob2}
  {\rm Pr}(H_2) = (p\phifix)^3 (1-p\phifix)^2 \chi^{(3)}_{m-2}(p,\phifix), \\
  \label{gcontlong.eq:allowableseqprob3}
  {\rm Pr}(H_3) = (p\phifix)^3 (1-p\phifix)^3 \chi^{(3)}_{m-3}(p,\phifix), \\
  \label{gcontlong.eq:allowableseqprob4}
  {\rm Pr}(H_4) = (p\phifix)^2 (1-p\phifix)^3 \chi^{(3)}_{m-2}(p,\phifix), \\
  \label{gcontlong.eq:allowableseqprob5}
  {\rm Pr}(H_5) = (p\phifix)^3 (1-p\phifix)^3 \chi^{(3)}_{m-3}(p,\phifix), \\
  \mbox{and} \nonumber \\
  \label{gcontlong.eq:allowableseqprob6}
  {\rm Pr}(H_6) = (p\phifix)^3 (1-p\phifix)^4 \chi^{(3)}_{m-4}(p,\phifix).
\end{gather}
Summing these will give us the probability $\Gamma_m$ 
but one small correction is needed for $m=1$.
By incorporating $d_{n+2}=0$ into the sequence 
of~\Req{gcontlong.eq:T3k2seq2b}, we considered only $m \ge 2$ sequences.
So we must also add in the probability of the $m=1$ sequence
given in~\Req{gcontlong.eq:T3k2seq2} which is $(p\phifix)^2 (1-p\phifix)^2$.
Combining this additional quantity with the probabilities in~\req{gcontlong.eq:allowableseqprob1}
through~\req{gcontlong.eq:allowableseqprob6}, we have
\begin{eqnarray}
  \label{gcontlong.eq:gammam}
  \lefteqn{\Gamma_m(p,\phifix,3) = (p\phifix)^2 (1-p\phifix)^2 \times } \\
  &  &  \left[ 
    \delta_{m1}
    + \chi^{(3)}_{m-1}
    + \chi^{(3)}_{m-2} \right. \nonumber \\
  &  & \left.
    + 2p\phifix (1-p\phifix)\chi^{(3)}_{m-3}
    + p\phifix (1-p\phifix)^2\chi^{(3)}_{m-4}
  \right], \nonumber
\end{eqnarray}
where $\delta_{ij}$ is the Kronecker delta function
and we have suppressed the dependencies of the $\chi^{(3)}_m$
on $p$ and $\phifix$.
Substituting this into equation~\Req{gcontlong.eq:Gamma},
we obtain \Req{gcontlong.eq:Gammafull}.

The calculation for $T=2$ follows along the same lines as above.
We now have $a=\{0\}$ and $b=\{1,0\}$ as our subsequences.
There is only one starting sequence,
\begin{equation}
  \label{gcontlong.eq:T2k2seq}
  \{d_{n-1},d_{n},d_{n+1}\} = \{1,1,0\},
\end{equation}
as well as one exceptional ending sequences, $\{d_{n+m}\} = \{1\}$.
Defining
\begin{equation}
  \label{gcontlong.eq:chiT}
  \chi^{(2)}_{m} = 
  \sum_{k=0}^{[m/2]}
  \binom{m-k}{k}
    (1-p\phifix)^{m-k}
    (p\phifix)^k,
\end{equation}
the probability of being above the threshold and then having no reinfections 
may be written as
\begin{equation}
  \label{gcontlong.eq:gammaT2k2}
  \Gamma_m(p,\phifix;2) = p^2 \phifix^2 ( 1 - p \phifix)
  \left[
    \chi_{m-1}^{(2)}
    + p \phifix \chi_{m-2}^{(2)}
    \right].
\end{equation}
Using~\Req{gcontlong.eq:gammaT2k2}
in \Req{gcontlong.eq:Gamma}
we obtain the form for $\Gamma$ 
given in \Req{gcontlong.eq:GammaT2k2}.

\section{Transition between class I and class II contagion models}
\label{gcontlong.app:transI-IIsec}

The transition between class I and class II models of
contagion occurs when a saddle-node bifurcation 
collides with the transcritical bifurcation lying on the $p$-axis.
To find this transition, we must determine when the slope
of the non-zero fixed point curve at the transcritical bifurcation
(i.e., at $p=p_c$ and $\phifix=0$) becomes infinite.
For the heterogeneous version of the model with $r=1$ and variable
$\{P_k\}$, we are able to determine the behavior near $p=p_c$ as follows.
We first rearrange the right hand side of \Req{gcontlong.eq:phifixgendosedstar} 
to obtain a polynomial in $p\phifix$:
\begin{eqnarray}
  \label{gcontlong.eq:transIcalc1}
  \lefteqn{\phifix 
    = 
    \sum_{k=1}^{T}
    \binom{T}{k}
    P_k
    (p\phifix)^{k}
    (1-p\phifix)^{T-k},
    }
  \nonumber \\
  & = & 
  \sum_{k=1}^{T}
  \binom{T}{k}
  P_k
  (p\phifix)^{k}
  \sum_{j=0}^{T-k}
  \binom{T-k}{j}
  (-p\phifix)^{j},
  \nonumber \\
  & = & 
  \sum_{k=1}^{T}
  \sum_{j=0}^{T-k}
  \binom{T}{k}
  \binom{T-k}{j}
  P_k
  (-1)^{j}(p\phifix)^{k+j},
  \nonumber \\
  & = & 
  \sum_{l=1}^{T}
  \sum_{k=1}^{m}
  \binom{T}{k}
  \binom{T-k}{m-k}
  P_k
  (-1)^{m-k}
  (p\phifix)^{m},
  \nonumber \\
  & = & 
  \sum_{m=1}^{T}
  C_m
  (p\phifix)^{m},
\end{eqnarray}
where we have changed the summation over $j$ and $k$
to one over $m=k+j$ and $k$.
The coefficients $C_m$ identified in the above
may be written more simply as
\begin{equation}
  \label{gcontlong.eq:ck}
  C_m = 
  (-1)^{m}
  \binom{T}{m}
  \sum_{k=1}^{m}
  (-1)^{k}
  \binom{m}{k}
  P_k,
\end{equation}
since 
\begin{eqnarray}
  \label{gcontlong.eq:ckcalc}  
    \binom{T}{k} \binom{T-k}{m-k}
    & = &
    \frac{T!}{k!(T-k)!}
    \frac{(T-k)!}{(m-k)!(T-l)!}
    \nonumber  \\
  & = & 
  \frac{T!}{m!(T-m)!}
  \frac{m!}{k!(l-k)!}
  \nonumber \\
  & = &
  \binom{T}{m} \binom{m}{k}.
\end{eqnarray}
Expanding \Req{gcontlong.eq:transIcalc1} to second order  about 
$p=p_c$ and $\phifix=0$, and writing $\tilde{p}=p-p_c$, we obtain
\begin{equation}
  \label{gcontlong.eq:transIcalc2}
  \phifix
  \simeq
  C_1 (\tilde{p}+p_c) \phifix
  + C_2 p_c^2 \phifix^2.
\end{equation}
From \Req{gcontlong.eq:ck}, $C_1 = T P_1 (= 1/p_c)$ 
and $C_2 = \binom{T}{2}(-2P_1 + P_2)$.  Using also
that $p_c = 1/(TP_1)$, \Req{gcontlong.eq:transIcalc2}
then yields
\begin{equation}
  \label{gcontlong.eq:transIcalc3}
  \phifix 
  \simeq
  \frac{C_1}{C_2 p_c^2} \tilde{p}
  =
  \frac{T^2 P_1^3}{(T-1)(P_1 - P_2/2)} \tilde{p}.
\end{equation}
Upon requiring $\dee{\phifix}/\dee{p}=\infty$,
the transition condition of \Req{gcontlong.eq:transIcond} follows.

If $C_2=0$ (i.e., $P_1=P_2/2$), the above calculation
is no longer valid and the system exhibits a continuous phase transition
with a nontrivial exponent at $p_c$.  More generally, when 
$C_2 = C_3 = \ldots = C_n = 0$, 
we find for small $\phifix$ and $\tilde{p}$ that
\begin{equation}
  \label{gcontlong.eq:exception}
  \phifix \simeq
  \frac{C_1}{C_{n+1} p_c^{n+1}} \tilde{p}^{1/n}.
\end{equation}
We observe that for $C_2 = C_3 = \ldots = C_n = 0$, 
a linearity condition in the $\{P_k\}$ must hold,
specifically, $P_k = k P_1$ for $k=1,\ldots,n$.
To show this, we rearrange the $\{C_m\}$ given
by \Req{gcontlong.eq:ck} as follows, ignoring
multiplicative factors and substituting $P_k=kP_1$:
\begin{eqnarray}
  \label{eq:gcontlong.ckmod}
  C_m & \propto & \sum_{k=1}^{m} (-1)^k \binom{m}{k} P_k
  =
  \sum_{k=1}^{m} (-1)^{k} \frac{m!}{(k-1)!(m-k)!} P_1,
  \nonumber \\
  & \propto & 
  \sum_{k=1}^{m} (-1)^{k} \frac{(m-1)!}{(k-1)!(m-k)!}
  \propto
  \sum_{k'=0}^{m-1} (-1)^{k'} \binom{m-1}{k'},
  \nonumber \\
  & = & [ 1 + (-1)]^{m-1} = 0,
\end{eqnarray}
where we have shifted the index $k$ to $k'=k-1$.

\end{document}